# Elastodynamic single-sided homogeneous Green's function representation: Theory and numerical examples

Christian Reinicke [*], Kees Wapenaar

*Delft University of Technology, Department of Geoscience and Engineering, Stevinweg 1, 2628 CN Delft, The Netherlands*

## HIGHLIGHTS

- The homogeneous Green's function: From a closed-towards an open-boundary representation.
- For propagating waves all orders of scattering as well as mode conversions are described correctly.
- Numerical analysis of limitations of the presented theory imposed by evanescent waves.
- Numerical example demonstrates evanescent wave tunnelling via the single-sided representation.

## ARTICLE INFO

## ABSTRACT

The homogeneous Green's function is the difference between an impulse response and its time-reversal. According to existing representation theorems, the homogeneous Green's function associated with source–receiver pairs inside a medium can be computed from measurements at a boundary enclosing the medium. However, in many applications such as seismic imaging, time-lapse monitoring, medical imaging, non-destructive testing, etc., media are only accessible from one side. A recent development of wave theory has provided a representation of the homogeneous Green's function in an elastic medium in terms of wavefield recordings at a single (open) boundary. Despite its single-sidedness, the elastodynamic homogeneous Green's function representation accounts for all orders of scattering inside the medium. We present the theory of the elastodynamic single-sided homogeneous Green's function representation and illustrate it with numerical examples for 2D laterally-invariant media. For propagating waves, the resulting homogeneous Green's functions match the exact ones within numerical precision, demonstrating the accuracy of the theory. In addition, we analyse the accuracy of the single-sided representation of the homogeneous Green's function for evanescent wave tunnelling.

## 1. Introduction

The homogeneous Green's function is the difference between an impulse response and its time-reversal. In the absence of losses, the wave equation is symmetric in time. Therefore, an impulse response to a source and its time-reversal satisfy the same wave equation. By subtracting the wave equations for these two responses from each other, we obtain a wave equation with a source term equal to zero, and a solution: the homogeneous Green's function.

---

[*] Corresponding author.
*E-mail addresses:* c.reinicke@tudelft.nl (C. Reinicke), c.p.a.wapenaar@tudelft.nl (K. Wapenaar).

In optics, Porter [1] used a closed-boundary representation of the homogeneous Green's function to retrieve the wavefield inside a medium. This representation has been the basis for inverse source problems [2] as well as inverse-scattering methods [3]. Unfortunately, in many practical situations, there is only single-sided access to the medium. When measurements are absent at a substantial part of the closed boundary, the retrieved homogeneous Green's function will suffer from significant artefacts. In particular in the presence of strong internal multiple scattering, these artefacts become more severe.

The closed-boundary representation can be modified to become an integral representation over the top and bottom boundaries of the medium if the medium has infinite horizontal extent [e.g. 4,5]. Further, a recent progress of wave theory has demonstrated that, after appropriate modification of the homogeneous Green's function representation, the integral contribution from the bottom boundary vanishes [6]. The result is a single-sided homogeneous Green's function representation. This representation correctly describes the wavefield inside the medium, including all orders of scattering, but excluding evanescent waves. The form of the single-sided representation is similar to the closed-boundary representation. However, the single-sided representation uses a so-called focusing function instead of the time-reversed Green's function [6]. For acoustic waves, the focusing function can be retrieved from a single-sided reflection response and an estimate of the direct arrival, using the Marchenko method [e.g. 7–9]. In the elastodynamic case, the approximate focusing function can be retrieved in a similar way [10,11]. However, an exact retrieval of the elastodynamic focusing function requires additional information about the medium [11]. In this paper, we assume that the elastodynamic focusing function is available (obtained either approximately by the Marchenko method or by direct modelling when the medium is known). The single-sided representation theorem provides the mathematical framework to place virtual sources and/or receivers inside the medium. Imaging techniques, e.g. for medical or geophysical applications with limited access to the medium, could benefit from this. Furthermore, virtual receivers inside a medium could be used for time-lapse monitoring purposes, i.e. to observe changes in a medium over time. Other potential applications could be forecasting the medium response to induced sources, or the localisation of passive sources inside a medium such as an earthquake [12].

We outline the theory of the single-sided homogeneous Green's function representation for elastodynamic waves in lossless media. Further, we evaluate the accuracy of the elastodynamic single-sided homogeneous Green's function representation numerically for 2D layered media. For the evaluation we use a directly modelled homogeneous Green's function as a reference. Eventually, we present an example in which the wavefield has wavenumber–frequency components that are evanescent only inside a thin layer between the recording boundary and the target depth. We demonstrate that these evanescent wavenumber–frequency components of the elastodynamic homogeneous Green's function are accounted for by the single-sided representation, except for small numerical errors.

## 2. Theory

This section consists of three parts. In the first part we show the definition of the decomposed matrix–vector wave equation. In the second part we define the elastodynamic homogeneous Green's function, and in the third part, we introduce the elastodynamic single-sided homogeneous Green's function representation. A detailed derivation of this representation can be found in Appendix A.

### 2.1. Matrix–vector wave equation for decomposed wavefields

We represent the elastodynamic wavefield using powerflux-normalised P- and S-wavefield potentials. Besides, we choose the depth direction $x_3$ as a preferential direction of propagation. For this reason, we decompose the wavefield in downward and upward propagating waves [13–15], and we define a $6 \times 1$ wave vector **p** containing decomposed wavefields and a $6 \times 1$ source vector **s** containing sources for these decomposed wavefields,

$$\mathbf{p} = \begin{pmatrix} \mathbf{p}^+ \\ \mathbf{p}^- \end{pmatrix}, \quad \mathbf{p}^+ = \begin{pmatrix} \Phi^+ \\ \Psi^+ \\ \Upsilon^+ \end{pmatrix}, \quad \mathbf{p}^- = \begin{pmatrix} \Phi^- \\ \Psi^- \\ \Upsilon^- \end{pmatrix}, \quad \mathbf{s} = \begin{pmatrix} \mathbf{s}^+ \\ \mathbf{s}^- \end{pmatrix}. \tag{1}$$

The superscript $+$ refers to downgoing waves, the superscript $-$ to upgoing waves. The wavefield potentials $\Phi^\pm$, $\Psi^\pm$, $\Upsilon^\pm$ represent the P, S1 and S2 waves, respectively (in cylindrical coordinates in a laterally-invariant medium, S1 and S2 waves are SV and SH waves). The decomposed source terms $\mathbf{s}^\pm$ are defined analogous to the decomposed wavefields $\mathbf{p}^\pm$. After applying a forward Fourier transform from the space–time to the space–frequency domain,

$$\mathbf{p}(\mathbf{x}, \omega) = \int \mathbf{p}(\mathbf{x}, t) e^{i\omega t} dt, \tag{2}$$

the matrix–vector wave equation for decomposed wavefields can be written as,

$$\partial_3 \mathbf{p}(\mathbf{x}, \omega) - \mathcal{B} \, \mathbf{p}(\mathbf{x}, \omega) = \mathbf{s}(\mathbf{x}, \omega). \tag{3}$$

Here, i is an imaginary unit ($i^2 = -1$), $\partial_3$ denotes a spatial derivative in the $x_3$ direction and the operator $\mathcal{B} = \mathcal{B}(\mathbf{x}, \omega)$ accounts for the propagation and the mutual coupling of the decomposed fields [an expression of $\mathcal{B}$ can be found in 6]. The spatial coordinates are denoted by $\mathbf{x} = (x_1, x_2, x_3)^T$, the time is denoted by $t$ and the frequency is denoted by $\omega$. The superscript $T$ denotes a transpose. Further details about decomposed wavefields in 3D inhomogeneous media can be found in [5,13,16–19].

## 2.2. Homogeneous Green's function

Consider the decomposed matrix–vector wave equation with a delta source term, $\mathbf{s} = \mathbf{I}\delta(\mathbf{x} - \mathbf{x}_s)$, where $\mathbf{I}$ is an identity matrix of appropriate size. The solution of this equation,

$$\partial_3 \boldsymbol{\Gamma}(\mathbf{x}, \mathbf{x}_s, \omega) - \mathcal{B}\, \boldsymbol{\Gamma}(\mathbf{x}, \mathbf{x}_s, \omega) = \mathbf{I}\delta(\mathbf{x} - \mathbf{x}_s), \tag{4}$$

is the Green's matrix $\boldsymbol{\Gamma}(\mathbf{x}, \mathbf{x}_s, \omega)$ containing decomposed wavefields,

$$\boldsymbol{\Gamma}(\mathbf{x}, \mathbf{x}_s, \omega) = \begin{pmatrix} \mathbf{G}^{+,+} & \mathbf{G}^{+,-} \\ \mathbf{G}^{-,+} & \mathbf{G}^{-,-} \end{pmatrix}(\mathbf{x}, \mathbf{x}_s, \omega). \tag{5}$$

Here, the first superscript describes the direction of the decomposed wavefields at the receiver position $\mathbf{x}$, the second superscript describes the direction of the decomposed source fields at the source position $\mathbf{x}_s$. The $3 \times 3$ submatrices $\mathbf{G}^{\pm,\pm}$ are defined as,

$$\mathbf{G}^{\pm,\pm}(\mathbf{x}, \mathbf{x}_s, \omega) = \begin{pmatrix} G^{\pm,\pm}_{\Phi,\Phi} & G^{\pm,\pm}_{\Phi,\Psi} & G^{\pm,\pm}_{\Phi,\Upsilon} \\ G^{\pm,\pm}_{\Psi,\Phi} & G^{\pm,\pm}_{\Psi,\Psi} & G^{\pm,\pm}_{\Psi,\Upsilon} \\ G^{\pm,\pm}_{\Upsilon,\Phi} & G^{\pm,\pm}_{\Upsilon,\Psi} & G^{\pm,\pm}_{\Upsilon,\Upsilon} \end{pmatrix}(\mathbf{x}, \mathbf{x}_s, \omega), \tag{6}$$

where the first subscript describes the wavefield potential at the receiver position $\mathbf{x}$ and the second subscript describes the wavefield potential at the source position $\mathbf{x}_s$. In this paper, we use $6 \times 6$ matrices to describe complete decomposed elastodynamic wavefields (e.g. in Eq. (5)) and $3 \times 3$ matrices to describe their four decomposed parts (e.g. in Eq. (6)). Ignoring evanescent waves, the operator $\mathcal{B}$ and its complex-conjugate $\mathcal{B}^*$ are mutually related as follows [combining Equations 70 and 88 in [20]],

$$\mathcal{B} = \mathbf{K}\mathcal{B}^*\mathbf{K}. \tag{7}$$

The superscript $*$ denotes a complex-conjugate. The matrix $\mathbf{K}$ as well as matrices $\mathbf{J}$ and $\mathbf{N}$ which we will use later are defined as,

$$\mathbf{K} = \begin{pmatrix} \mathbf{O} & \mathbf{I} \\ \mathbf{I} & \mathbf{O} \end{pmatrix}, \quad \mathbf{J} = \begin{pmatrix} \mathbf{I} & \mathbf{O} \\ \mathbf{O} & -\mathbf{I} \end{pmatrix}, \quad \mathbf{N} = \begin{pmatrix} \mathbf{O} & \mathbf{I} \\ -\mathbf{I} & \mathbf{O} \end{pmatrix}, \tag{8}$$

where $\mathbf{O}$ is a null matrix of appropriate size. The matrices $\mathbf{K}$ and $\mathbf{J}$ can be thought of as a Pauli matrices, the matrix $\mathbf{N}$ is a symplectic matrix.

We complex-conjugate Eq. (4) and substitute the operator $\mathcal{B}^*$ using Eq. (7). By pre- and post-multiplying the result by the matrix $\mathbf{K}$, we obtain,

$$\partial_3 \left[\mathbf{K}\boldsymbol{\Gamma}^*(\mathbf{x}, \mathbf{x}_s, \omega)\mathbf{K}\right] - \mathcal{B}\left[\mathbf{K}\boldsymbol{\Gamma}^*(\mathbf{x}, \mathbf{x}_s, \omega)\mathbf{K}\right] = \mathbf{I}\delta(\mathbf{x} - \mathbf{x}_s). \tag{9}$$

Here, we used the identity $\mathbf{KK} = \mathbf{I}$. According to this equation, the quantity $\mathbf{K}\boldsymbol{\Gamma}^*(\mathbf{x}, \mathbf{x}_s, \omega)\mathbf{K}$ is another solution of the wave equation. It is the Fourier transform of a time-reversed Green's function, but the diagonal as well as the off-diagonal elements $\mathbf{G}^{\pm,\pm}$ are interchanged. By subtracting Eq. (9) from Eq. (4) we obtain the homogeneous wave equation, i.e. a wave equation with a source term equal to zero,

$$\partial_3 \boldsymbol{\Gamma}_h(\mathbf{x}, \mathbf{x}_s, \omega) - \mathcal{B}\, \boldsymbol{\Gamma}_h(\mathbf{x}, \mathbf{x}_s, \omega) = \mathbf{O}. \tag{10}$$

A solution of the homogeneous wave equation is the homogeneous Green's function,

$$\boldsymbol{\Gamma}_h(\mathbf{x}, \mathbf{x}_s, \omega) = \boldsymbol{\Gamma}(\mathbf{x}, \mathbf{x}_s, \omega) - \mathbf{K}\boldsymbol{\Gamma}^*(\mathbf{x}, \mathbf{x}_s, \omega)\mathbf{K}, \tag{11}$$

which contains block-matrices as follows,

$$\boldsymbol{\Gamma}_h(\mathbf{x}, \mathbf{x}_s, \omega) = \begin{pmatrix} \mathbf{G}_h^{+,+} & \mathbf{G}_h^{+,-} \\ \mathbf{G}_h^{-,+} & \mathbf{G}_h^{-,-} \end{pmatrix}(\mathbf{x}, \mathbf{x}_s, \omega) = \begin{pmatrix} \mathbf{G}^{+,+} - (\mathbf{G}^{-,-})^* & \mathbf{G}^{+,-} - (\mathbf{G}^{-,+})^* \\ \mathbf{G}^{-,+} - (\mathbf{G}^{+,-})^* & \mathbf{G}^{-,-} - (\mathbf{G}^{+,+})^* \end{pmatrix}(\mathbf{x}, \mathbf{x}_s, \omega). \tag{12}$$

Eq. (11) states that, in the space–frequency domain, the homogeneous Green's function is the Green's function minus its complex-conjugate, pre- and post multiplied by matrix $\mathbf{K}$. From the source–receiver reciprocity relation of the decomposed Green's function [6],

$$\boldsymbol{\Gamma}(\mathbf{x}, \mathbf{x}_s, \omega) = \mathbf{N}\boldsymbol{\Gamma}^T(\mathbf{x}_s, \mathbf{x}, \omega)\mathbf{N}, \tag{13}$$

and the identity $\mathbf{KN} = -\mathbf{NK}$, it follows that the homogeneous Green's function obeys the source–receiver reciprocity relation,

$$\boldsymbol{\Gamma}_h(\mathbf{x}, \mathbf{x}_s, \omega) = \mathbf{N}\boldsymbol{\Gamma}_h^T(\mathbf{x}_s, \mathbf{x}, \omega)\mathbf{N}. \tag{14}$$

*2.3. Elastodynamic single-sided homogeneous Green's function representation*

Consider a medium which is bounded by a reflection-free boundary $\partial \mathbb{D}_0$ at the top ($x_3 = x_{3,0}$) as depicted in Fig. 1(a). Let $\mathbf{R}(\mathbf{x}, \mathbf{x}', \omega)$ be the reflection response associated with a source for downgoing waves located at $\mathbf{x}'$ just above $\partial \mathbb{D}_0$, and a receiver for upgoing waves located at $\mathbf{x}$ on $\partial \mathbb{D}_0$. We define the depth level of $\mathbf{x}'$ as $\partial \mathbb{D}'_0$. In the notation of decomposed wavefields, we can state that $\mathbf{R}(\mathbf{x}, \mathbf{x}', \omega) = \mathbf{G}^{-,+}(\mathbf{x}, \mathbf{x}', \omega)$. The direct (downgoing) wave from $\mathbf{x}'$ to $\mathbf{x}$ on $\partial \mathbb{D}_0$ is part of the decomposed Green's function $\mathbf{G}^{+,+}(\mathbf{x}, \mathbf{x}', \omega) = \mathbf{I}\delta(\mathbf{x}_H - \mathbf{x}'_H)$. Here, the subscript $H$ refers to the horizontal components, i.e. $\mathbf{x}_H = (x_1, x_2)^T$. Since the medium is reflection-free above $\partial \mathbb{D}_0$, the decomposed Green's functions associated with sources for upgoing waves at $\mathbf{x}'$ and receivers at $\mathbf{x}$ on $\partial \mathbb{D}_0$ are zero, $\mathbf{G}^{\pm,-}(\mathbf{x}, \mathbf{x}', \omega) = \mathbf{O}$. According to the matrix notation in Eq. (12), we can write the homogeneous Green's matrix $\mathbf{\Gamma}_h(\mathbf{x}, \mathbf{x}', \omega)$ for $\mathbf{x}$ at $\partial \mathbb{D}_0$ in terms of the reflection response $\mathbf{R}(\mathbf{x}, \mathbf{x}', \omega)$ and an identity matrix $\mathbf{I}$ of appropriate size,

$$\mathbf{\Gamma}_h(\mathbf{x}, \mathbf{x}', \omega) = \begin{pmatrix} \mathbf{I}\delta(\mathbf{x}_H - \mathbf{x}'_H) & -\mathbf{R}^*(\mathbf{x}, \mathbf{x}', \omega) \\ \mathbf{R}(\mathbf{x}, \mathbf{x}', \omega) & -\mathbf{I}\delta(\mathbf{x}_H - \mathbf{x}'_H) \end{pmatrix}. \tag{15}$$

Next, we introduce the focusing function $\mathbf{F}(\mathbf{x}, \mathbf{x}_s, \omega)$. The focusing function is defined in a so-called truncated medium which is identical to the true medium for $x_{3,0} \leq x_3 < x_{3,s}$ and homogeneous for $x_3 \geq x_{3,s}$, where $x_{3,s}$ is the depth of the focal point at $\mathbf{x}_s$ (see Fig. 1(b)). We assume that $\mathbf{x}_s$ is located below $\partial \mathbb{D}_0$. The decomposed focusing matrix consists of a down- and an upgoing part,

$$\mathbf{F}(\mathbf{x}, \mathbf{x}_s, \omega) = \begin{pmatrix} \mathbf{F}^+ \\ \mathbf{F}^- \end{pmatrix}(\mathbf{x}, \mathbf{x}_s, \omega), \tag{16}$$

with,

$$\mathbf{F}^\pm(\mathbf{x}, \mathbf{x}_s, \omega) = \begin{pmatrix} F^\pm_{\Phi,\Phi} & F^\pm_{\Phi,\Psi} & F^\pm_{\Phi,\Upsilon} \\ F^\pm_{\Psi,\Phi} & F^\pm_{\Psi,\Psi} & F^\pm_{\Psi,\Upsilon} \\ F^\pm_{\Upsilon,\Phi} & F^\pm_{\Upsilon,\Psi} & F^\pm_{\Upsilon,\Upsilon} \end{pmatrix}(\mathbf{x}, \mathbf{x}_s, \omega). \tag{17}$$

The superscripts $\pm$ and the first subscript describe the wavefield direction and the wavefield potential at the receiver location $\mathbf{x}$, respectively. The second subscript describes the wavefield potential at the focusing position $\mathbf{x}_s$. The downgoing focusing function $\mathbf{F}^+(\mathbf{x}, \mathbf{x}_s, \omega)$ for $\mathbf{x}$ at $\partial \mathbb{D}_0$ is the inverse of a transmission response of the truncated medium between $\partial \mathbb{D}_0$ and the depth level $x_3 = x_{3,s}$ [6],

$$\int_{\partial \mathbb{D}_0} \mathbf{T}^+(\mathbf{x}, \mathbf{x}', \omega) \mathbf{F}^+(\mathbf{x}', \mathbf{x}_s, \omega) \mathrm{d}^2 \mathbf{x}'_H \bigg|_{x_3 = x_{3,s}} = \mathbf{I}\, \delta(\mathbf{x}_H - \mathbf{x}_{H,s}), \tag{18}$$

and the complete focusing function $\mathbf{F}$ obeys the focusing condition,

$$\mathbf{F}(\mathbf{x}, \mathbf{x}_s, \omega)|_{x_3 = x_{3,s}} = \mathbf{I}_1\, \delta(\mathbf{x}_H - \mathbf{x}_{H,s}). \tag{19}$$

Here, we introduced the matrix $\mathbf{I}_1 = (\mathbf{I}, \mathbf{O})^T$. The upgoing focusing function $\mathbf{F}^-(\mathbf{x}, \mathbf{x}_s, \omega)$ is the reflection response of the downgoing focusing function in the truncated medium. In a physical interpretation, the focusing function is the Fourier transform of a wavefield injected at the surface $\partial \mathbb{D}_0$, which focuses at time zero at the position $\mathbf{x}_s$ (see Fig. 1(b)). Note, that the first event of the focusing function is injected at negative times.

Consider the homogeneous Green's function $\mathbf{\Gamma}_h(\mathbf{x}, \mathbf{x}_s, \omega)$ related to a source at $\mathbf{x}_s$ inside the medium and receivers on the surface $\partial \mathbb{D}_0$ as depicted in Fig. 2(a). According to Eqs. (A.17) and (A.18) in Appendix A, the homogeneous Green's function $\mathbf{\Gamma}_h(\mathbf{x}, \mathbf{x}', \omega)$ recorded on the top boundary and the focusing function $\mathbf{F}(\mathbf{x}, \mathbf{x}_s, \omega)$ can construct a wavefield $\mathbf{\Gamma}_1(\mathbf{x}, \mathbf{x}_s, \omega)$,

$$\mathbf{\Gamma}_1(\mathbf{x}, \mathbf{x}_s, \omega) = \int_{\partial \mathbb{D}'_0} \mathbf{\Gamma}_h(\mathbf{x}, \mathbf{x}', \omega) \mathbf{F}(\mathbf{x}', \mathbf{x}_s, \omega) \mathbf{I}_1^T \mathrm{d}^2 \mathbf{x}', \tag{20}$$

from which the homogeneous Green's function $\mathbf{\Gamma}_h(\mathbf{x}, \mathbf{x}_s, \omega)$ can be represented as,

$$\mathbf{\Gamma}_h(\mathbf{x}, \mathbf{x}_s, \omega) = \mathbf{\Gamma}_1(\mathbf{x}, \mathbf{x}_s, \omega) - \mathbf{K}\mathbf{\Gamma}_1^*(\mathbf{x}, \mathbf{x}_s, \omega)\mathbf{K}. \tag{21}$$

Evaluating the matrix products in Eqs. (20)–(21) reveals that the forward-in-time propagating part of the homogeneous Green's function $\mathbf{\Gamma}_h(\mathbf{x}, \mathbf{x}_s, \omega)$ is a superposition of the non-zero sub-matrices of $\mathbf{\Gamma}_1(\mathbf{x}, \mathbf{x}_s, \omega)$, i.e. $\mathbf{G}^{-,+}(\mathbf{x}, \mathbf{x}_s, \omega)$ and $\mathbf{G}^{-,-}(\mathbf{x}, \mathbf{x}_s, \omega)$. Hence, Fig. 2(a) also illustrates the wavefield $\mathbf{\Gamma}_1(\mathbf{x}, \mathbf{x}_s, \omega)$.

The homogeneous Green's function $\mathbf{\Gamma}_h(\mathbf{x}, \mathbf{x}', \omega)$ corresponds to sources and receivers at the surface $(\partial \mathbb{D}_0 \cup \partial \mathbb{D}'_0)$. A physical interpretation of Eq. (20) is that the focusing function focuses, or inverse-propagates, the sources of $\mathbf{\Gamma}_h(\mathbf{x}, \mathbf{x}', \omega)$ from $\mathbf{x}'$ to $\mathbf{x}_s$.

In analogy, according to Eqs. (A.12) and (A.13) in Appendix A, the receivers of the homogeneous Green's function $\mathbf{\Gamma}_h(\mathbf{x}, \mathbf{x}_s, \omega)$ are focused on, or inverse-propagated to, a virtual receiver location $\mathbf{x}_r$ inside the medium, according to,

$$\mathbf{\Gamma}_2(\mathbf{x}_r, \mathbf{x}_s, \omega) = \int_{\partial \mathbb{D}_0} \mathbf{I}_2 \mathbf{F}^T(\mathbf{x}, \mathbf{x}_r, \omega) \mathbf{N} \mathbf{\Gamma}_h(\mathbf{x}, \mathbf{x}_s, \omega) \mathrm{d}^2 \mathbf{x}, \tag{22}$$

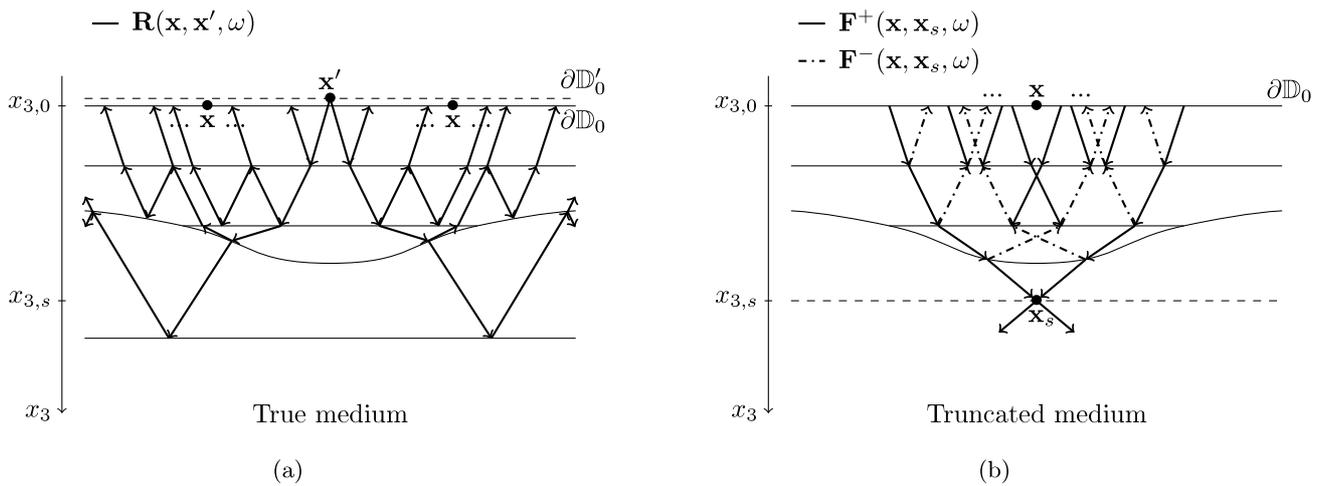

**Fig. 1.** Reflection response and focusing function. (a) $\mathbf{R}(\mathbf{x}, \mathbf{x}', \omega)$ is the reflection response of a medium which is reflection-free above $\partial \mathbb{D}_0$. The source is located at $\mathbf{x}'$ on $\partial \mathbb{D}_0'$ (just above $\partial \mathbb{D}_0$), the receivers are located at $\mathbf{x}$ on $\partial \mathbb{D}_0$. (b) The decomposed focusing function $\mathbf{F}(\mathbf{x}, \mathbf{x}_s, \omega)$ is defined in a truncated medium which is identical to the true medium for $x_3 < x_{3,s}$, and reflection-free for $x_3 \geq x_{3,s}$.

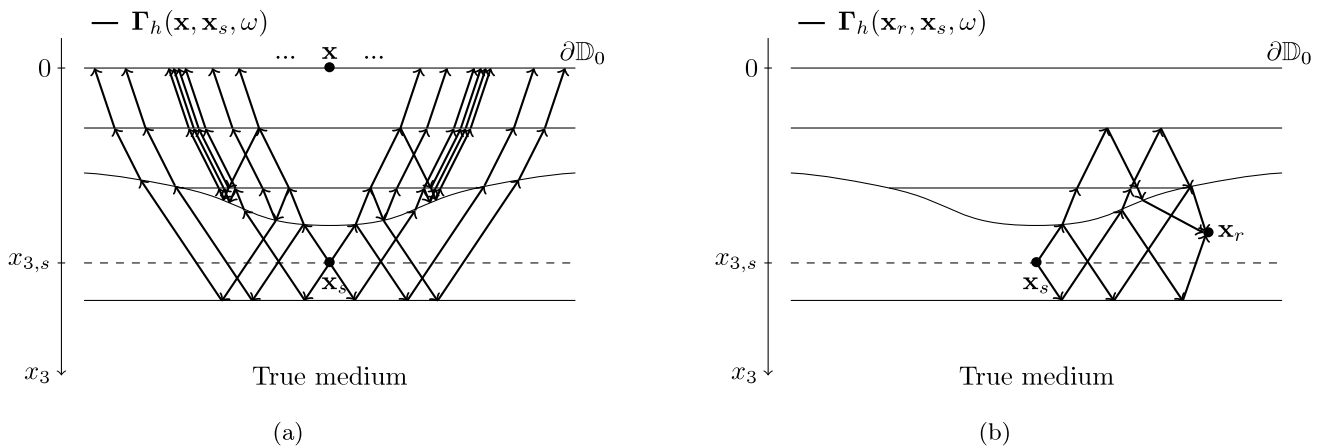

**Fig. 2.** Homogeneous Green's functions. (a) $\boldsymbol{\Gamma}_h(\mathbf{x}, \mathbf{x}_s, \omega)$ is the homogeneous Green's function related to a source located inside the medium at $\mathbf{x}_s$ and receivers located on the surface $\partial \mathbb{D}_0$ at $\mathbf{x}$. (b) $\boldsymbol{\Gamma}_h(\mathbf{x}_r, \mathbf{x}_s, \omega)$ is the homogeneous Green's function related to a source located inside the medium at $\mathbf{x}_s$ and a receiver located inside the medium at $\mathbf{x}_r$. Both subfigures show the forward-in-time propagating part of the respective homogeneous Green's function.

from which the homogeneous Green's function $\boldsymbol{\Gamma}_h(\mathbf{x}_r, \mathbf{x}_s, \omega)$ can be constructed,

$$\boldsymbol{\Gamma}_h(\mathbf{x}_r, \mathbf{x}_s, \omega) = \boldsymbol{\Gamma}_2(\mathbf{x}_r, \mathbf{x}_s, \omega) - \mathbf{K}\boldsymbol{\Gamma}_2^*(\mathbf{x}_r, \mathbf{x}_s, \omega)\mathbf{K}. \tag{23}$$

The quantity $\mathbf{F}(\mathbf{x}, \mathbf{x}_r, \omega)$ is the focusing function related to a focal point at $\mathbf{x}_r$. Further, we introduced the matrix $\mathbf{I}_2 = (\mathbf{O}, \mathbf{I})^T$. Eqs. (20)–(23) together form an elastodynamic single-sided homogeneous Green's function representation: It expresses the homogeneous Green's function between $\mathbf{x}_s$ and $\mathbf{x}_r$ inside the medium in terms of the single-sided data $\boldsymbol{\Gamma}_h(\mathbf{x}, \mathbf{x}', \omega)$ at the upper boundary $\partial \mathbb{D}_0 \cup \partial \mathbb{D}_0'$. An illustration of the homogeneous Green's function $\boldsymbol{\Gamma}_h(\mathbf{x}_r, \mathbf{x}_s, \omega)$ is displayed in Fig. 2(b). By evaluating the matrix products in Eqs. (22)–(23) it can be seen that the forward-in-time propagating part of the wavefield $\boldsymbol{\Gamma}_2(\mathbf{x}_r, \mathbf{x}_s, \omega)$ is represented by all paths in Fig. 2(b) that are associated with upward propagating waves at $\mathbf{x}_r$. The representation formed by Eqs. (20)–(23) involves integrations along an open boundary of the medium, and therefore, only requires single-sided access to the medium.

## 3. Numerical example

In this section, we show a numerical example of the elastodynamic single-sided homogeneous Green's function representation for a 2D laterally-invariant medium. Further, we investigate the accuracy of the single-sided representation for wavenumber–frequency components of the elastodynamic homogeneous Green's function that are evanescent only inside a thin layer between the recording boundary and the virtual receiver depth. From here on, we consider a 2D medium, i.e. in all expressions of Section 2 the spatial coordinate $\mathbf{x}$ simplifies to $\mathbf{x} = (x_1, x_3)^T$ and the horizontal coordinate $\mathbf{x}_H$ simplifies to $x_1$. Besides, we only consider P and SV waves, indicated by the subscripts $\Phi$ and $\Psi$, respectively. In the provided numerical examples, we use modelled focusing functions. This allows us to analyse the properties of the single-sided representation, independent of approximations of the focusing function retrieval via the Marchenko method.

*3.1. Wavenumber–frequency domain expressions*

Since we consider a laterally-invariant model the required data can be modelled efficiently by wavefield extrapolation in the wavenumber–frequency domain [21,22]. However, in the theory section the single-sided homogeneous Green's function representation is formulated in the space–frequency domain. To transform the expressions to the wavenumber–frequency domain, we use the fact that all the presented expressions have a similar form, namely a product of two space-dependent functions $\mathbf{g}(x_1, x_3, x_1'', x_3'')$ and $\mathbf{f}(x_1'', x_3'', x_1', x_3')$, integrated across a horizontal surface $\partial\mathbb{D}''$,

$$\int_{\partial\mathbb{D}''} \mathbf{g}(x_1, x_3, x_1'', x_3'') \mathbf{f}(x_1'', x_3'', x_1', x_3') dx_1''. \tag{24}$$

In laterally-invariant media, integrals of the above form can be rewritten as a spatial convolution,

$$\int_{\partial\mathbb{D}''} \mathbf{g}(x_1 - x_1'', x_3, 0, x_3'') \mathbf{f}(x_1'', x_3'', x_1', x_3') dx_1'', \tag{25}$$

which corresponds to a multiplication in the wavenumber domain [23],

$$\tilde{\mathbf{g}}(k_1, x_3, 0, x_3'') \tilde{\mathbf{f}}(k_1, x_3'', x_1', x_3'). \tag{26}$$

Here, we introduced the horizontal wavenumber $k_1$. Note, when we say wavenumber domain, we refer to the horizontal-wavenumber domain $(k_1, x_3)$ on the receiver side (first coordinate). Wavefields in the wavenumber domain are written with a tilde on top. Expressions in the space and wavenumber domain are mutually related via the Fourier transform,

$$\tilde{\mathbf{g}}(k_1, x_3, x_1'', x_3'') = \int \mathbf{g}(x_1, x_3, x_1'', x_3'') e^{-ik_1 x_1} dx_1. \tag{27}$$

We model the required input data, the reflection response $\tilde{\mathbf{R}}(k_1, x_{3,0}, 0, x_{3,0}', \omega)$ and the focusing functions $\tilde{\mathbf{F}}(k_1, x_{3,0}, 0, x_{3,s/r}, \omega)$, in the wavenumber–frequency domain by wavefield extrapolation. Since we model all fields for a source with a horizontal space coordinate $x_1 = 0$, we will omit this coordinate in the following expressions. Next, we transform Eqs. (20)–(23) to the wavenumber domain according to Eqs. (25)–(27). After evaluating Eqs. (20)–(23) in the wavenumber–frequency domain, we transform the final result $\tilde{\Gamma}_h(k_1, x_{3,r}, x_{3,s}, \omega)$ to the space–time domain via an inverse Fourier transform,

$$\Gamma_h(\mathbf{x}_r, \mathbf{x}_s, t) = \frac{1}{(2\pi)^2} \iint \tilde{\Gamma}_h(k_1, x_{3,r}, x_{3,s}, \omega) e^{-i[\omega t - k_1(x_{1,r} - x_{1,s})]} dk_1 d\omega. \tag{28}$$

Here, we replaced the horizontal receiver coordinate $x_{1,r}$ in the exponent by the horizontal offset between the receiver and the source, $x_{1,r} - x_{1,s}$, to account for the actual horizontal position of the source $x_{1,s}$.

*3.2. Results*

We present a numerical example of the elastodynamic single-sided homogeneous Green's function representation. The result is compared to the exact homogeneous Green's function, which is computed by wavefield extrapolation. For a clear illustration, we choose a simple model as depicted in Fig. 3. Results for a more complex model can be found in Appendix B. Note that we use superscripts $(i)$ to refer to the $i$th layer of the medium.

According to Appendix A, the single-sided homogeneous Green's function representation ignores evanescent waves, which are associated with imaginary-valued vertical-wavenumbers $k_3(c_{p/s}, k_1, \omega) = i\sqrt{k_1^2 - \frac{\omega^2}{c_{p/s}^2}}$. Thus, all wavenumber–frequency components that satisfy the relation,

$$|k_1| > \frac{\omega}{c_{max}}, \tag{29}$$

should be excluded from the analysis. Here, $c_{max}$ is the maximum propagation velocity of the truncated medium.

In the following, before displaying a wavefield in the space–time domain, we mute evanescent waves according to Eq. (29) using the maximum P-wave velocity of the truncated medium $c_{max}$, apply an inverse Fourier transform according to Eq. (28) and convolve the resulting wavefield with a 30 Hz Ricker wavelet [defined by Eq. 7 in 24]. Note that, we create a virtual source as well as a virtual receiver, meaning that there are two truncated media, bounded at the bottom by $x_{3,s}$ and $x_{3,r}$, respectively. The above mentioned maximum P-wave velocity $c_{max}$ is the overall maximum P-wave velocity of both truncated media.

The reflection response $\tilde{\mathbf{R}}$ and the focusing functions $\tilde{\mathbf{F}}^\pm$ are modelled by wavefield extrapolation using the modelling parameters shown in Table 1.

| | | | |
|---|---|---|---|
| $x_{3,0} = 0$ m | $c_p^{(1)} = 1500$ m s$^{-1}$ | $c_s^{(1)} = 800$ m s$^{-1}$ | $\rho^{(1)} = 1000$ kg m$^{-3}$ |
| $x_3 = 500$ m | $c_p^{(1)} = 1500$ m s$^{-1}$ | $c_s^{(1)} = 800$ m s$^{-1}$ | $\rho^{(1)} = 1000$ kg m$^{-3}$ |
| $x_3 = 1250$ m | $c_p^{(2)} = 2000$ m s$^{-1}$ | $c_s^{(2)} = 1000$ m s$^{-1}$ | $\rho^{(2)} = 1500$ kg m$^{-3}$ |
| $x_{3,s} = 1500$ m | $c_p^{(3)} = 2500$ m s$^{-1}$ | $c_s^{(3)} = 1200$ m s$^{-1}$ | $\rho^{(3)} = 1000$ kg m$^{-3}$ |
| $x_3 = 2000$ m | $c_p^{(3)} = 2500$ m s$^{-1}$ | $c_s^{(3)} = 1200$ m s$^{-1}$ | $\rho^{(3)} = 1000$ kg m$^{-3}$ |
| $x_3 = 2500$ m | $c_p^{(4)} = 3000$ m s$^{-1}$ | $c_s^{(4)} = 1400$ m s$^{-1}$ | $\rho^{(4)} = 1500$ kg m$^{-3}$ |
| $x_3 = 3000$ m | $c_p^{(5)} = 3500$ m s$^{-1}$ | $c_s^{(5)} = 1600$ m s$^{-1}$ | $\rho^{(5)} = 1000$ kg m$^{-3}$ |
| | $c_p^{(5)} = 3500$ m s$^{-1}$ | $c_s^{(5)} = 1600$ m s$^{-1}$ | $\rho^{(5)} = 1000$ kg m$^{-3}$ |

**Fig. 3.** Layered model. The model depth ranges from 0 m to 3000 m, the lateral distance ranges from $-12\,812.5$ m to 12 800 m. The P-wave velocity, S-wave velocity and density are denoted by $c_p^{(i)}$, $c_s^{(i)}$ and $\rho^{(i)}$ respectively. The superscripts ($i$) refer to the $i$th layer of the medium. The top boundary and the virtual source depth are denoted by $x_{3,0}$ and $x_{3,s}$, respectively. Solid lines represent medium interfaces and dashed lines represent (scattering-free) depth levels.

**Table 1**
Modelling parameters.

| | |
|---|---|
| Number of frequency samples | 1025 |
| Frequency sample interval | 0.511 s$^{-1}$ |
| Number of wavenumber samples | 2048 |
| Wavenumber sample interval | $0.245 \times 10^{-3}$ m$^{-1}$ |

The modelled reflection response $\tilde{\mathbf{R}}(k_1, x_{3,0}, x'_{3,0}, \omega)$ contains P-wave and S-wave recordings. The components associated with a P-wave source are superimposed and shown in the space–time domain in Fig. 4(a). For clearer illustration, we only show source–receiver offsets between $-2000$ m and $+2000$ m and times between 0 s and 4 s. The reflection response contains primary reflections, converted reflections and internal multiples. Mode conversions between P- and S-waves do not occur at zero-incidence. Therefore, events with a gap at $x_1 = 0$ m are easily identified as converted waves.

Next, we define a virtual source inside the medium at $\mathbf{x}_s = (0$ m, $1500$ m$)^T$. Thus, we model a focusing function $\tilde{\mathbf{F}}(k_1, x_{3,0}, x_{3,s}, \omega)$ with focusing depth equal to 1500 m. The up- and downgoing P- and S-wave components of the focusing function, that are associated with a P-wave focus, are superimposed and displayed in the space–time domain in Fig. 4(b). Since the focusing function contains both causal and acausal events it is displayed for times between $-2$ s and $+2$ s.

The reflection response and focusing function of Figs. 4(a) and 4(b) are used to compute a homogeneous Green's function $\tilde{\Gamma}_h(k_1, x_{3,0}, x_{3,s}, \omega)$ related to the virtual source at $\mathbf{x}_s$ and receivers at the surface. The computation is performed as stated in Eqs. (20)–(21). Fig. 4(c) shows a superposition of the up- and downgoing P- and S-wave components of the resulting homogeneous Green's function in the space–time domain. Only responses to a virtual P-wave source are displayed. Fig. 4(c) illustrates that the acausal part of the homogeneous Green's function is a time-reversed and polarity-flipped version of the causal part.

Subsequently, we compute the single-sided representation of the homogeneous Green's function $\tilde{\Gamma}_h(k_1, x_{3,r}, x_{3,s}, \omega)$ associated with virtual receivers at $x_{3,r} = 1700$ m according to Eqs. (22)–(23). We superpose the downgoing components of the homogeneous Green's function, $\tilde{\mathbf{G}}_h^{+,+}(k_1, x_{3,r}, x_{3,s}, \omega) + \tilde{\mathbf{G}}_h^{+,-}(k_1, x_{3,r}, x_{3,s}, \omega)$, and display the absolute value of the result in Fig. 5. Here, we display the four elastic components separately. Due to the symmetry of the homogeneous Green's function $\tilde{\Gamma}_h(k_1, x_{3,r}, x_{3,s}, \omega)$ (see Eq. (12)), the sum of its upgoing components, $\tilde{\mathbf{G}}_h^{-,+}(k_1, x_{3,r}, x_{3,s}, \omega) + \tilde{\mathbf{G}}_h^{-,-}(k_1, x_{3,r}, x_{3,s}, \omega)$, produces an identical result. Further, since the medium is horizontally-layered we only show positive wavenumbers $k_1$. In Fig. 5, the amplitude of the $\Phi\Phi$, $\Phi\Psi$ and $\Psi\Phi$ components of the quantity, $\tilde{\mathbf{G}}_h^{+,+}(k_1, x_{3,r}, x_{3,s}, \omega) + \tilde{\mathbf{G}}_h^{+,-}(k_1, x_{3,r}, x_{3,s}, \omega)$, decreases rapidly for $|k_1| > \frac{\omega}{c_p^{(3)}}$, i.e. beyond the dashed red line. The velocity $c_p^{(3)}$ is the maximum propagation velocity in the truncated medium, and therefore, the line $|k_1| = \frac{\omega}{c_p^{(3)}}$ separates propagating from evanescent waves. As shown in Appendix A, the elastodynamic single-sided homogeneous Green's function representation does not take into account wavenumber–frequency components that are evanescent on the boundaries of the domain (here $x_3 = x_{3,0}$ and $x_3 = x_{3,r}$). Thus, for wavenumbers $|k_1| > \frac{\omega}{c_p^{(3)}}$ the retrieved quantity, $\tilde{\mathbf{G}}_h^{+,+}(k_1, x_{3,r}, x_{3,s}, \omega) + \tilde{\mathbf{G}}_h^{+,-}(k_1, x_{3,r}, x_{3,s}, \omega)$, is not representing the analytic elastodynamic homogeneous Green's function. Further, in the wavenumber regime $|k_1| \gg \frac{\omega}{c_p^{(3)}}$, the quantity, $\tilde{\mathbf{G}}_h^{+,+}(k_1, x_{3,r}, x_{3,s}, \omega) + \tilde{\mathbf{G}}_h^{+,-}(k_1, x_{3,r}, x_{3,s}, \omega)$, becomes unstable. This instability could be due to either the behaviour of the elastodynamic single-sided homogeneous Green's function representation for evanescent wavenumber–frequency components, or numerical instabilities, or both. Nevertheless, the analytic elastodynamic homogeneous Green's function is characterised by an exponential amplitude decay for evanescent wavenumber–frequency components. The amplitude of the $\Psi\Psi$ component of the quantity, $\tilde{\mathbf{G}}_h^{+,+}(k_1, x_{3,r}, x_{3,s}, \omega) + \tilde{\mathbf{G}}_h^{+,-}(k_1, x_{3,r}, x_{3,s}, \omega)$, is in the order of one for $|k_1| < \frac{\omega}{c_p^{(2)}}$ and increases rapidly for $|k_1| > \frac{\omega}{c_p^{(2)}}$, beyond the indicated dotted red line. Hence, for the $\Psi\Psi$ component of the quantity, $\tilde{\mathbf{G}}_h^{+,+}(k_1, x_{3,r}, x_{3,s}, \omega) + \tilde{\mathbf{G}}_h^{+,-}(k_1, x_{3,r}, x_{3,s}, \omega)$, the transition from propagating to evanescent waves is defined by the P-wave velocity of the second layer $c_p^{(2)}$, instead of the P-wave velocity of the third layer $c_p^{(3)}$. This is expected because its $\Psi\Psi$ component only requires an S-wave focus in the third layer of the medium. Creating an S-wave focus in the third layer only allows for P-wave conversions above layer (3). As such, the highest P-wave velocity associated with an S-wave focus in layer (3) is $c_p^{(2)}$. In other words, the unstable behaviour of the $\Psi\Psi$ component for $|k_1| > \frac{\omega}{c_p^{(2)}}$ (to the right of the dotted red line) is caused by waves that are S-waves at the source, convert to P-waves in the second layer and convert back to S-waves before reaching the receivers. For unconverted S-waves the highest propagation velocity is $c_s^{(3)}$, i.e. those waves are propagating for $|k_1| < \frac{\omega}{c_s^{(3)}}$ (to the left of the dashed blue line). However, these propagating S-waves are obscured by the unstable behaviour of the parts of the $\Psi\Psi$ component that convert to P-waves in the second layer.

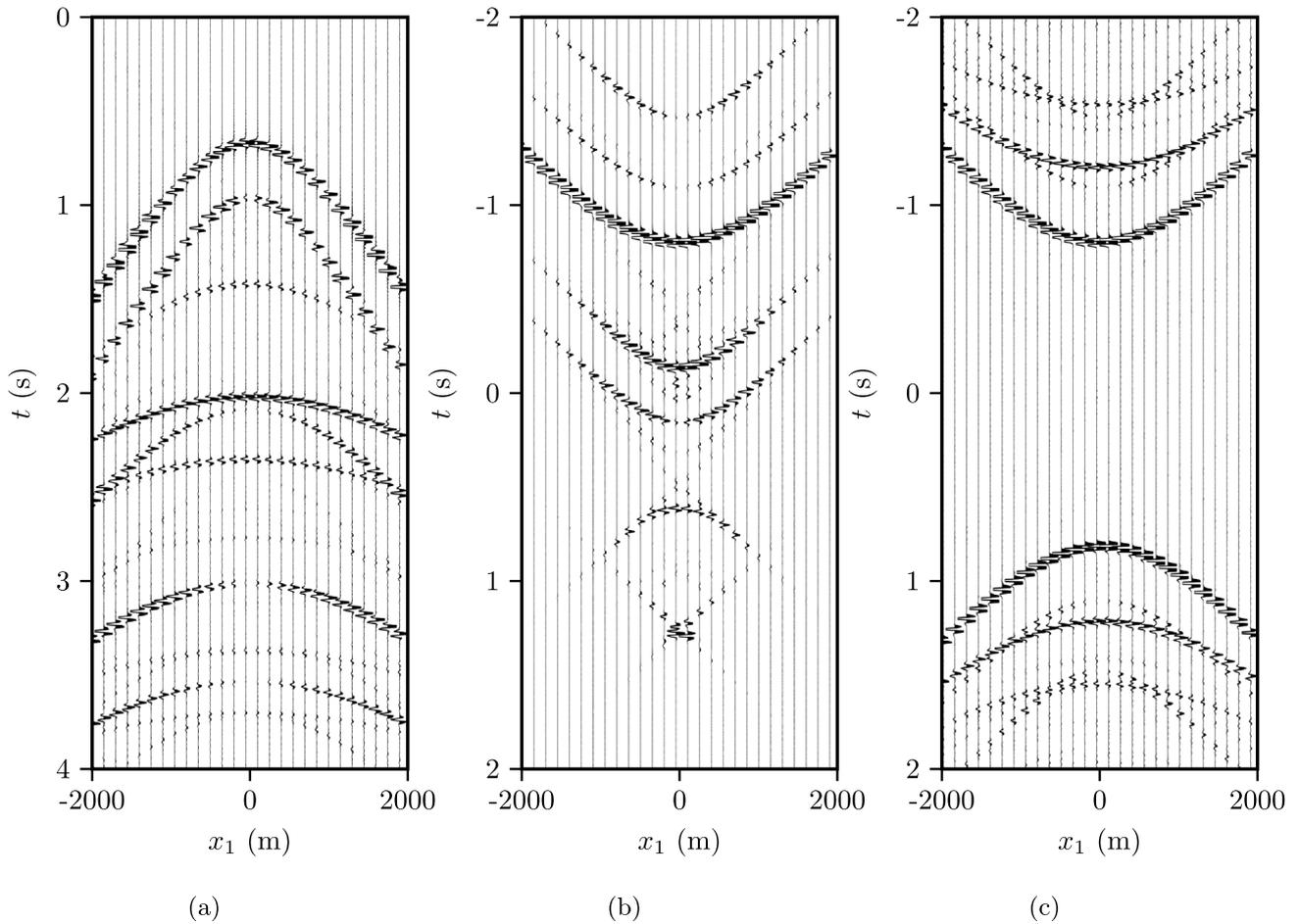

**Fig. 4.** Virtual source retrieval. (a) Reflection response $\mathbf{R}(x_1, x_{3,0}, x'_{3,0}, t)$. (b) Focusing function $\mathbf{F}(x_1, x_{3,0}, x_{3,s}, t)$ for a focal point at $\mathbf{x}_s = (0\,\mathrm{m}, 1500\,\mathrm{m})^T$. (c) Homogeneous Green's function $\Gamma_h(x_1, x_{3,0}, x_{3,s}, t)$, obtained from the single-sided representation, for a virtual source at $\mathbf{x}_s = (0\,\mathrm{m}, 1500\,\mathrm{m})^T$ and receivers at the surface. Fields in (a) and (c) are associated with a P-wave source, and the field in (b) is associated with a P-wave focus. A $k_1$–$\omega$ filter was applied to all displayed fields before transformation to the space–time domain. The traces in Figures (a) and (c) were multiplied with a gain function ($\times\,e^{0.5|t|}$) to emphasise the late arrivals. Note that the plots have different clipping.

We evaluate the accuracy of the elastodynamic single-sided homogeneous Green's function representation by comparing it to the exact elastodynamic homogeneous Green's function. To that end, we model the elastodynamic homogeneous Green's function $\tilde{\Gamma}_{h,mod}(k_1, x_{3,r}, x_{3,s}, \omega)$ for an actual source at $\mathbf{x}_s = (0\,\mathrm{m}, 1500\,\mathrm{m})^T$. Next, we compute the relative error

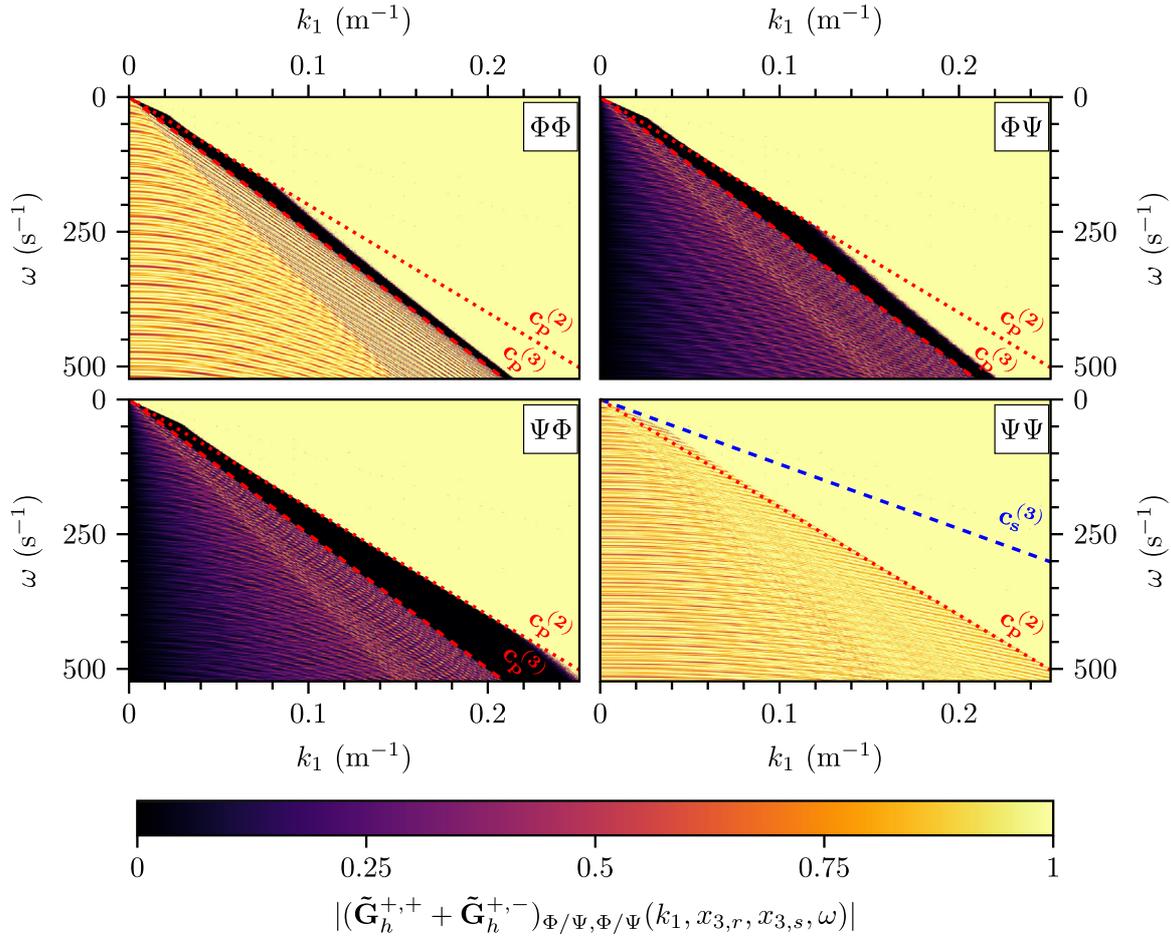

**Fig. 5.** Analysis of the retrieved elastodynamic homogeneous Green's function. The four figures show the single-sided representation of the elastodynamic homogeneous Green's function after summing its receiver-side downgoing components and taking the absolute value of the result $|\tilde{\mathbf{G}}_h^{+,+}(k_1, x_{3,r}, x_{3,s}, \omega) + \tilde{\mathbf{G}}_h^{+,-}(k_1, x_{3,r}, x_{3,s}, \omega)|$. The elastodynamic homogeneous Green's function is associated with a virtual source at $\mathbf{x}_s = (0\,\text{m}, 1500\,\text{m})^T$ and virtual receivers at $x_{3,r} = 1700$ m. We only show positive wavenumbers $k_1$. In addition, we draw lines, $k_1 = \frac{\omega}{c_{p/s}^{(i)}}$, defined by the P-/S-wave velocity $c_{p/s}^{(i)}$ in the $i$th layer of the model. The amplitudes in the plain yellow areas increase rapidly and were clipped for values greater than one. (For interpretation of the references to colour in this figure legend, the reader is referred to the web version of this article.)

of the single-sided representation of the elastodynamic homogeneous Green's function according to,

$$\tilde{\mathbf{E}}(k_1, x_{3,r}, x_{3,s}, \omega) = \frac{|(\tilde{\mathbf{G}}_h^{+,+} + \tilde{\mathbf{G}}_h^{+,-} - \tilde{\mathbf{G}}_{h,mod}^{+,+} - \tilde{\mathbf{G}}_{h,mod}^{+,-})(k_1, x_{3,r}, x_{3,s}, \omega)|}{|(\tilde{\mathbf{G}}_{h,mod}^{+,+} + \tilde{\mathbf{G}}_{h,mod}^{+,-})(k_1, x_{3,r}, x_{3,s}, \omega)|}, \quad (30)$$

where the absolute value is evaluated element-wise. The resulting relative error $\tilde{\mathbf{E}}(k_1, x_{3,r}, x_{3,s}, \omega)$ is shown in Fig. 6. For $|k_1| < \frac{\omega}{c_p^{(3)}}$, the single-sided representation of the quantity, $\tilde{\mathbf{G}}_h^{+,+}(k_1, x_{3,r}, x_{3,s}, \omega) + \tilde{\mathbf{G}}_h^{+,-}(k_1, x_{3,r}, x_{3,s}, \omega)$, is accurate within a relative error $\tilde{\mathbf{E}}(k_1, x_{3,r}, x_{3,s}, \omega)$ of about $10^{-15}$, i.e. within numerical precision. For $|k_1| > \frac{\omega}{c_p^{(3)}}$, the relative error $\tilde{\mathbf{E}}(k_1, x_{3,r}, x_{3,s}, \omega)$ increases drastically, except for the $\Psi\Psi$ component $\tilde{E}_{\Psi\Psi}(k_1, x_{3,r}, x_{3,s}, \omega)$. As explained above, the single-sided representation of the $\Psi\Psi$ component of the homogeneous Green's function $\tilde{\Gamma}_{h,\Psi\Psi}(k_1, x_{3,r}, x_{3,s}, \omega)$ only requires an S-wave focus in the third layer of the medium, such that the $\Psi\Psi$ component of the quantity, $\tilde{\mathbf{G}}_h^{+,+}(k_1, x_{3,r}, x_{3,s}, \omega) + \tilde{\mathbf{G}}_h^{+,-}(k_1, x_{3,r}, x_{3,s}, \omega)$, is accurate within numerical precision up to wavenumbers defined by the P-wave velocity of the second layer of the medium, $|k_1| < \frac{\omega}{c_p^{(2)}}$.

The resulting elastodynamic homogeneous Green's function $\tilde{\Gamma}_h(k_1, x_{3,r}, x_{3,s}, \omega)$ is a decomposed wavefield, described by up- and downgoing P- and S-waves. To obtain the full homogeneous Green's function $\tilde{\mathbf{G}}_h(k_1, x_{3,r}, x_{3,s}, \omega)$, that is associated with measurable wavefield quantities, we apply wavefield composition,

$$\tilde{\mathbf{G}}_h(k_1, x_{3,r}, x_{3,s}, \omega) = \tilde{\mathcal{L}}(k_1, x_{3,r}, \omega)\tilde{\Gamma}_h(k_1, x_{3,r}, x_{3,s}, \omega)\tilde{\mathcal{L}}^{-1}(k_1, x_{3,s}, \omega), \quad (31)$$

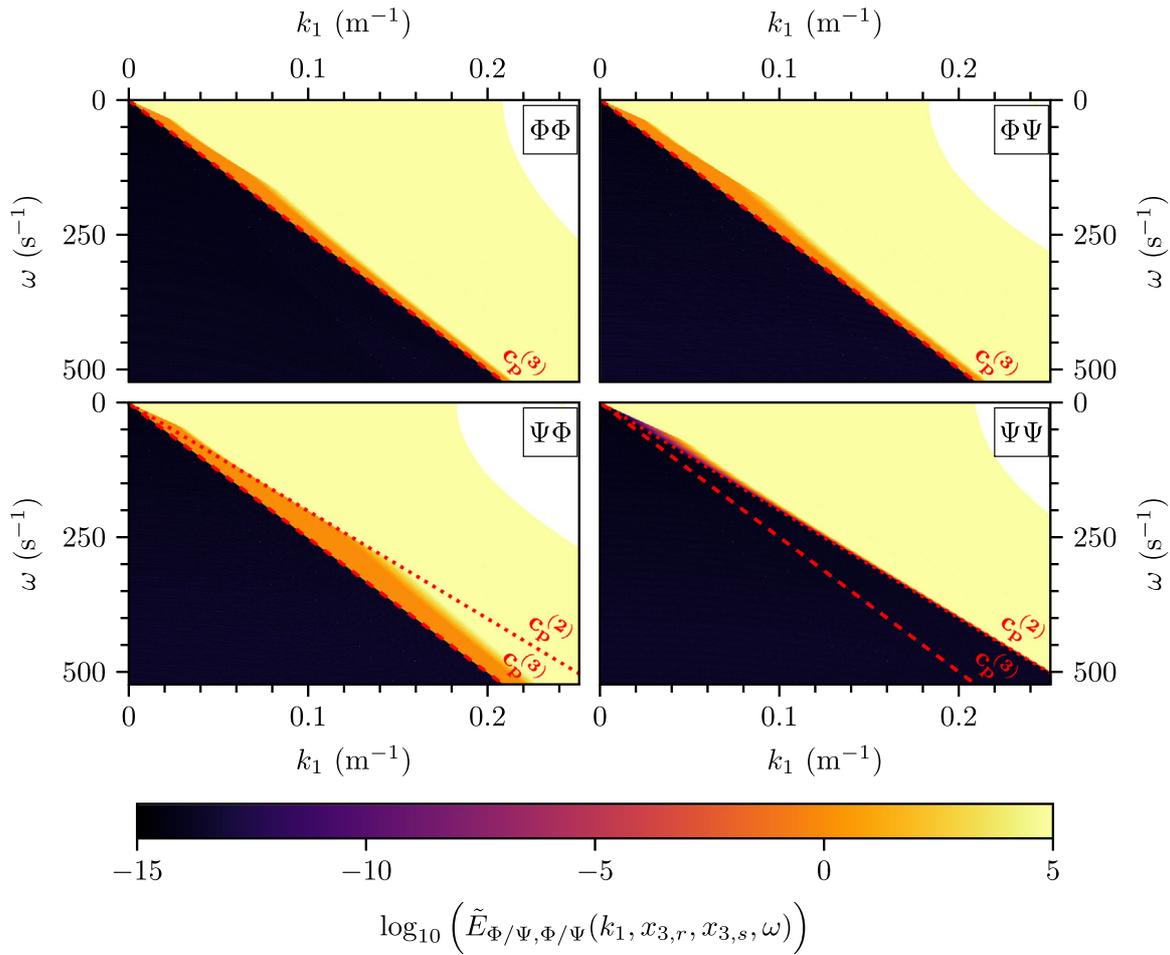

**Fig. 6.** Relative error of the retrieved elastodynamic homogeneous Green's function. The four figures show the relative error defined by Eq. (30). The red lines are $k_1 = \frac{\omega}{c_{p/s}^{(i)}}$ defined by the P-wave velocity $c_p^{(i)}$ in the $i$th layer of the model. The amplitudes in the plain yellow area increase rapidly and were clipped for values greater than five. Inside the white area the relative error could not be represented as a number due to limited numerical precision (double-precision). Of course, these values are still defined but their representation requires a higher numerical precision. (For interpretation of the references to colour in this figure legend, the reader is referred to the web version of this article.)

where $\tilde{\mathcal{L}}(k_1, x_{3,r/s}, \omega)$ is the composition operator. Further details about wavefield composition can be found in Wapenaar et al. [6]. The full homogeneous Green's function $\tilde{\mathbf{G}}_h(k_1, x_{3,r}, x_{3,s}, \omega)$ relates force sources $\mathbf{f}$ and deformation sources $\mathbf{h}$ to traction recordings $\boldsymbol{\tau}$ and particle velocity recordings $\mathbf{v}$,

$$\tilde{\mathbf{G}}_h(k_1, x_{3,r}, x_{3,s}, \omega) = \begin{pmatrix} \tilde{\mathbf{G}}^{\tau,\mathbf{f}} - (\tilde{\mathbf{G}}^{\tau,\mathbf{f}})^* & \tilde{\mathbf{G}}^{\tau,\mathbf{h}} + (\tilde{\mathbf{G}}^{\tau,\mathbf{h}})^* \\ \tilde{\mathbf{G}}^{\mathbf{v},\mathbf{f}} + (\tilde{\mathbf{G}}^{\mathbf{v},\mathbf{f}})^* & \tilde{\mathbf{G}}^{\mathbf{v},\mathbf{h}} - (\tilde{\mathbf{G}}^{\mathbf{v},\mathbf{h}})^* \end{pmatrix} (k_1, x_{3,r}, x_{3,s}, \omega). \tag{32}$$

From Eq. (32) follows that, in the space–time domain, the $(\boldsymbol{\tau}, \mathbf{f})$ and $(\mathbf{v}, \mathbf{h})$ components of the full homogeneous Green's function are anti-symmetric in time, and hence, vanish at time zero. This is undesirable for imaging applications because imaging uses the wavefield at time zero for $\mathbf{x}_r = \mathbf{x}_s$. However, in the space–time domain the $(\boldsymbol{\tau}, \mathbf{h})$ and $(\mathbf{v}, \mathbf{f})$ components are non-zero at time zero for $\mathbf{x}_r = \mathbf{x}_s$, and can be used for imaging.

After transforming the elastodynamic homogeneous Green's function to the space–time domain $\mathbf{G}_h(\mathbf{x}_r, \mathbf{x}_s, t)$, we display its $(v_3, f_3)$ component, $G^{v_3, f_3}(\mathbf{x}_r, \mathbf{x}_s, t) + G^{v_3, f_3}(\mathbf{x}_r, \mathbf{x}_s, -t)$, in Fig. 7. The time slices illustrate the symmetry of the homogeneous Green's function $\mathbf{G}_h^{\mathbf{v},\mathbf{f}}$ in time. At time zero the wavefield focuses. The focus is distorted by linear artefacts. The artefacts occur because the homogeneous Green's function was filtered by a $k_1$–$\omega$ mask that is determined by the maximum propagation velocity at a given virtual receiver depth $x_{3,r}$. The $k_1$–$\omega$ mask mutes the part of the propagating S-wavefield that overlaps with the evanescent P-wavefield, causing linear artefacts in the space–time domain. According to Snell's law, the inclination angle $\alpha$ of these linear artefacts is determined by,

$$\frac{1}{c_p} = \frac{\sin(\alpha)}{c_s}. \tag{33}$$

Using $c_p^{(3)} = 2500 \text{ m s}^{-1}$ and $c_s^{(3)} = 1200 \text{ m s}^{-1}$, we find that the linear artefact at the focusing position at time zero has an inclination angle $\alpha = 28.7°$ (see Fig. 7). An inspection of the time slices, e.g. for $t = 0.15$ s and $t = 0.45$ s, demonstrates that the single-sided representation of the elastodynamic homogeneous Green's function contains primary waves, converted waves, and multiply scattered waves.

We evaluate the accuracy of the elastodynamic single-sided homogeneous Green's function representation by comparing it to the exact elastodynamic homogeneous Green's function. To that end, we modelled the elastodynamic homogeneous Green's function for an actual source at $\mathbf{x}_s = (0 \text{ m}, 1500 \text{ m})^T$. The modelled elastodynamic homogeneous Green's function is subtracted from the elastodynamic homogeneous Green's function obtained from the single-sided representation. To exclude evanescent waves, the result $\Delta \tilde{\Gamma}_h(k_1, x_{3,r}, x_{3,s}, \omega)$ is element-wise multiplied by a $k_1$–$\omega$ filter $\tilde{\mathbf{M}}$ determined by the maximum propagation velocity of the medium. Next, we compute the normalised Frobenius norm $N_F = \frac{1}{\sqrt{4nt\,4nr}} \|\tilde{\mathbf{M}} \circ \Delta \tilde{\Gamma}_h(k_1, x_{3,r}, x_{3,s}, \omega)\|_2$ at each virtual receiver depth level $x_{3,r}$, where the symbol $\circ$ denotes a Hadamard product. The normalisation factor is a function of the number of time samples $nt$ and the number of space samples $nr$. We choose the normalisation factor $\sqrt{4nt\,4nr}$ because $\Delta \tilde{\Gamma}_h(k_1, x_{3,r}, x_{3,s}, \omega)$ consists of four $nt \times nr$ block-matrices. In Fig. 8, we show the resulting residual norm as a function of virtual receiver depth $x_{3,r}$. The difference plot demonstrates that, for all modelled wavenumber–frequency components of the propagating wavefield, the elastodynamic single-sided homogeneous Green's function representation is accurate within numerical precision. For the evaluation, we used the residual of the elastodynamic decomposed homogeneous Green's function $\Delta \tilde{\Gamma}_h(k_1, x_{3,r}, x_{3,s}, \omega)$ instead of its composed equivalent to exclude effects of the wavefield composition (see Eq. (31)) from the analysis. In this case, however, the wavefield composition also performs within numerical precision.

### 3.3. Evanescent wave tunnelling

In this section, we investigate tunnelling of evanescent waves using a new model (see Fig. 9). The new model is nearly identical to the model in Fig. 3. However, the layer between $x_3 = 1250$ m and $x_3 = 2000$ m is replaced by a thin layer ranging from $x_3 = 1250$ m to $x_3 = 1330$ m and a thicker layer ranging from $x_3 = 1330$ m to $x_3 = 2000$ m. For the thin layer, the P- and S-wave velocities are $2500 \text{ m s}^{-1}$ and $1200 \text{ m s}^{-1}$, respectively. For the layer below, the propagation velocities are smaller, $c_p^{(4)} = 2000 \text{ m s}^{-1}$ and $c_s^{(4)} = 1000 \text{ m s}^{-1}$.

Next, we evaluate the single-sided representation of the elastodynamic homogeneous Green's function $\tilde{\Gamma}_h(k_1, x_{3,r}, x_{3,s}, \omega)$ according to Eqs. (20)–(23), for a virtual source at $x_{3,s} = 1500$ m and a virtual receiver at $x_{3,r} = 1700$ m. To analyse the result, we model the elastodynamic homogeneous Green's function $\tilde{\Gamma}_{h,mod}(k_1, x_{3,r}, x_{3,s}, \omega)$ as a reference and compute the relative error $\tilde{\mathbf{E}}(k_1, x_{3,r}, x_{3,s}, \omega)$ according to Eq. (30). Fig. 10 shows the resulting relative error. For wavenumber–frequency components that are only propagating in the truncated medium, i.e. $|k_1| < \frac{\omega}{c_p^{(3)}}$, the relative error $\tilde{\mathbf{E}}(k_1, x_{3,r}, x_{3,s}, \omega)$ is in the order of $10^{-15}$, as expected. For larger wavenumbers, $|k_1| > \frac{\omega}{c_p^{(3)}}$, we would expect a rapid increase of the relative error $\tilde{\mathbf{E}}(k_1, x_{3,r}, x_{3,s}, \omega)$ due to instabilities, similar to Fig. 6. However, Fig. 10 shows that for wavenumbers, $\frac{\omega}{c_p^{(3)}} < |k_1| < \frac{\omega}{c_p^{(2)}}$, the relative error $\tilde{\mathbf{E}}(k_1, x_{3,r}, x_{3,s}, \omega)$ ranges from about $10^{-15}$ to about $10^{-6}$. The relative error is not within numerical precision but still very small. We interpret this effect as evanescent wave tunnelling. In the theory section, we excluded evanescent waves. However, this restriction is only required on the boundaries on which the reciprocity theorem of the correlation-type is evaluated (see Appendix A). Therefore, the theory does not exclude evanescent wave tunnelling. Besides, the theory does not make any assumption about the thickness of the tunnelled layer. In the presented numerical example, we observed that the relative error of the tunnelling experiment depends on the thickness $\Delta x_3$ of the tunnelled layer. This observation is expected because the amplitude of the tunnelled evanescent waves is reduced by the factor,

$$\exp\left(-\sqrt{k_1^2 - \frac{\omega^2}{c^2}} \Delta x_3\right). \tag{34}$$

The single-sided representation compensates for this amplitude decay. For increasing thickness of the tunnelled layer, i.e. stronger amplitude decay, the amplitude ratio of tunnelled and propagating waves becomes smaller, and the numerical errors become larger. To estimate the maximum amplitude decay in the presented example, we maximise $\sqrt{k_1^2 - \frac{\omega^2}{c^2}}$ (see Eq. (34)) inside the wavenumber regime, $\frac{\omega}{c_p^{(3)}} < |k_1| < \frac{\omega}{c_p^{(2)}}$. Thus, we choose $c = c_p^{(3)}$, maximise the frequency $\omega = 1025 \times 0.511 \text{ s}^{-1}$ (see Table 1) and maximise the horizontal wavenumber $k_1 = \frac{\omega}{c_p^{(2)}}$. The resulting amplitude decay factor,

$$\exp\left(-\sqrt{k_1^2 - \frac{\omega^2}{c^2}} \Delta x_3\right) = \exp\left(-\sqrt{\left(\frac{1025 \times 0.511 \text{ s}^{-1}}{2000 \text{ m s}^{-1}}\right)^2 - \left(\frac{1025 \times 0.511 \text{ s}^{-1}}{2500 \text{ m s}^{-1}}\right)^2} \times 80.0 \text{ m}\right)$$

$$= \exp\left(-0.157 \text{ m}^{-1} \times 80.0 \text{ m}\right) = 3.51 \times 10^{-6}, \tag{35}$$

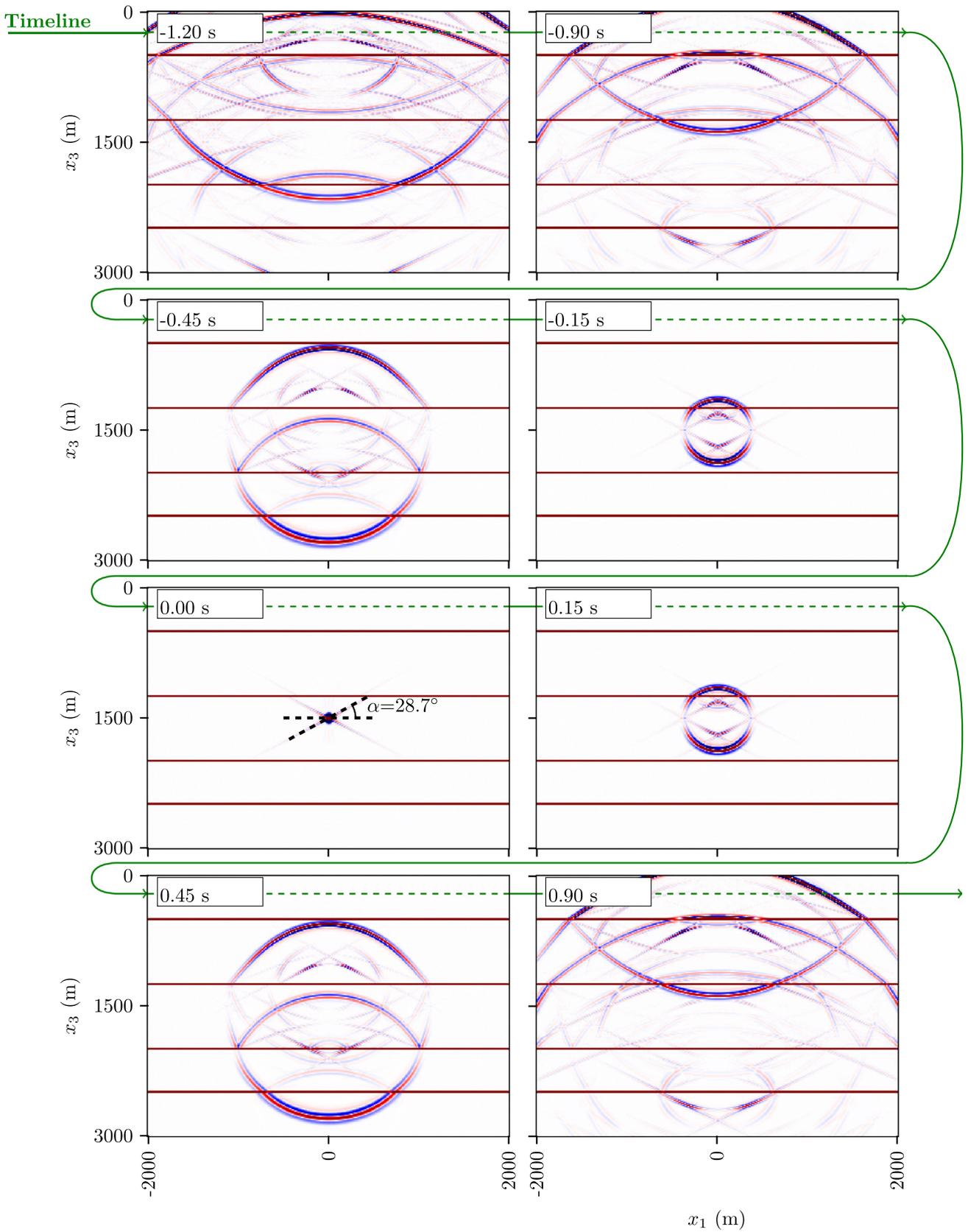

**Fig. 7.** Single-sided representation of the elastodynamic homogeneous Green's function. The time slices show the single-sided representation of the elastodynamic homogeneous Green's function $\mathbf{G}_h^{v_3,f_3}(\mathbf{x}_r, \mathbf{x}_s, t)$ related to a virtual source ($f_3$) at $\mathbf{x}_s = (0\,\text{m}, 1500\,\text{m})^T$ and virtual receivers ($v_3$) placed on a grid with a depth range from 0 m to 3000 m and a lateral distance range from −2000 m to 2000 m. The spatial sampling interval is 12.5 m in both horizontal and vertical directions. The time slices were multiplied by a gain function ($\times e^{1.5|t|}$) to emphasise the late arrivals. At time zero, we indicate the slope angle $\alpha$ of a linear artefact caused by applying a $k_1$–$\omega$ mask.

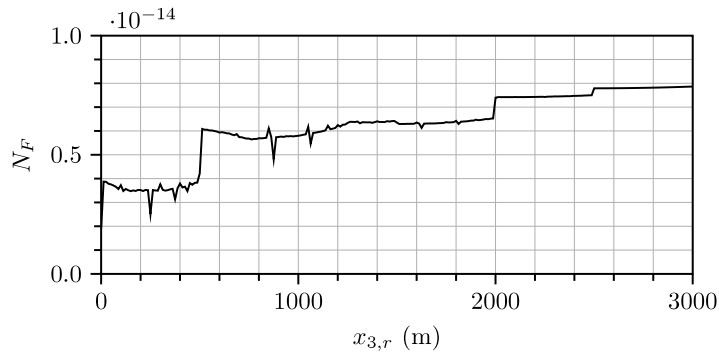

**Fig. 8.** Error analysis. Normalised Frobenius norm $N_F$ of the difference $\Delta \tilde{\Gamma}_h(k_1, x_{3,r}, x_{3,s}, \omega)$ between the elastodynamic single-sided homogeneous Green's function representation (see Fig. 7) and its modelled equivalent as a function of virtual receiver depth $x_{3,r}$.

| | | | |
|---|---|---|---|
| $x_{3,0} = 0$ m | $c_p^{(1)} = 1500$ m s$^{-1}$ | $c_s^{(1)} = 800$ m s$^{-1}$ | $\rho^{(1)} = 1000$ kg m$^{-3}$ |
| $x_3 = 500$ m | $c_p^{(1)} = 1500$ m s$^{-1}$ | $c_s^{(1)} = 800$ m s$^{-1}$ | $\rho^{(1)} = 1000$ kg m$^{-3}$ |
| $x_3 = 1250$ m | $c_p^{(2)} = 2000$ m s$^{-1}$ | $c_s^{(2)} = 1000$ m s$^{-1}$ | $\rho^{(2)} = 1500$ kg m$^{-3}$ |
| $x_3 = 1330$ m | $c_p^{(3)} = 2500$ m s$^{-1}$ | $c_s^{(3)} = 1200$ m s$^{-1}$ | $\rho^{(3)} = 1000$ kg m$^{-3}$ |
| $x_{3,s} = 1500$ m | $c_p^{(4)} = 2000$ m s$^{-1}$ | $c_s^{(4)} = 1000$ m s$^{-1}$ | $\rho^{(4)} = 1000$ kg m$^{-3}$ |
| $x_3 = 2000$ m | $c_p^{(4)} = 2000$ m s$^{-1}$ | $c_s^{(4)} = 1000$ m s$^{-1}$ | $\rho^{(4)} = 1000$ kg m$^{-3}$ |
| $x_3 = 2500$ m | $c_p^{(5)} = 3000$ m s$^{-1}$ | $c_s^{(5)} = 1400$ m s$^{-1}$ | $\rho^{(5)} = 1500$ kg m$^{-3}$ |
| $x_3 = 3000$ m | $c_p^{(6)} = 3500$ m s$^{-1}$ | $c_s^{(6)} = 1600$ m s$^{-1}$ | $\rho^{(6)} = 1000$ kg m$^{-3}$ |
| | $c_p^{(6)} = 3500$ m s$^{-1}$ | $c_s^{(6)} = 1600$ m s$^{-1}$ | $\rho^{(6)} = 1000$ kg m$^{-3}$ |

**Fig. 9.** Layered model. As Fig. 3 but with an additional thin layer between $x_3 = 1250$ m and $x_3 = 1300$ m. For the sake of readability, the distances between the interfaces are not proportional because the third layer is only 80 m thick.

shows that, for tunnelling through the high-velocity layer of 80.0 m thickness, the smallest amplitude ratio of tunnelled and propagating waves is already in the order of $10^{-6}$.

## 4. Discussion and conclusion

### 4.1. Discussion

We presented numerical examples of the single-sided homogeneous Green's function representation in elastic laterally-invariant media. Such 1D models are relatively simple, but they allow us to model the required fields with very high numerical accuracy via wavefield extrapolation. Nonetheless, the theory is valid for laterally varying media, and has already been tested numerically for acoustic waves [25]. In the future, we will extend the here presented numerical example to elastic laterally varying media.

Here, we modelled the focusing functions. In practice, for the acoustic situation the focusing functions can be retrieved from the reflection response combined with a smooth velocity model via the Marchenko method, which is described by Broggini et al. [26], Van der Neut et al. [27] and others. For example, Wapenaar et al. [25] presented an acoustic single-sided homogeneous Green's function representation that uses focusing functions retrieved via the Marchenko method. Da Costa Filho et al. [10] and Wapenaar [11] extended the Marchenko method to elastodynamic waves, however, it still requires more prior knowledge of the medium than in the acoustic case. In the here presented numerical example, we used numerically modelled elastodynamic focusing functions to analyse the properties of the elastodynamic single-sided homogeneous Green's function representation, independent of approximations of the focusing functions. Currently, we are developing the (elastodynamic) Marchenko method further to minimise the required amount of prior knowledge of the medium.

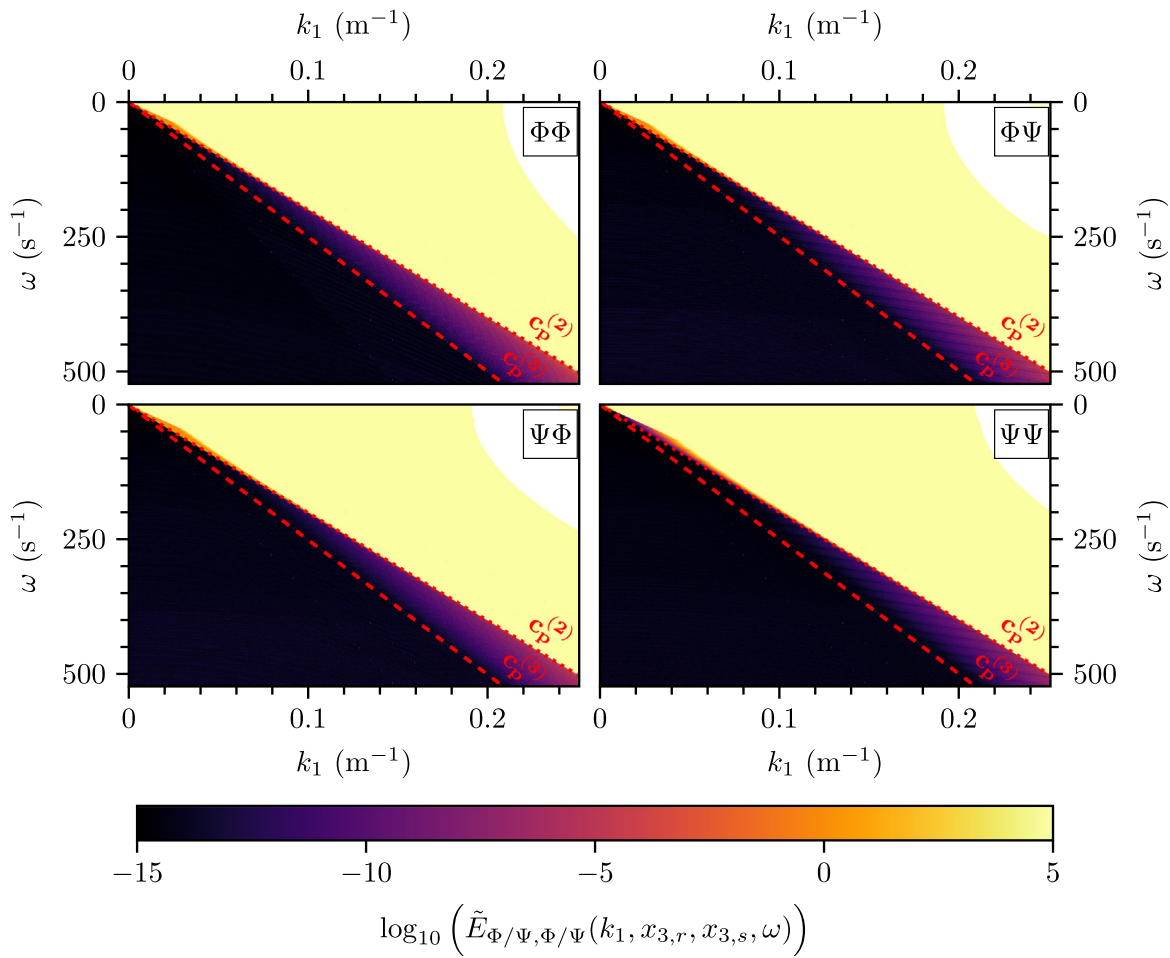

**Fig. 10.** Relative error of the retrieved elastodynamic homogeneous Green's function. As Fig. 6, but this time for the model in Fig. 9 with the thin high-velocity layer. Note that also for tunnelled waves (the region between the dashed and dotted red lines) the relative errors are very small.

According to the theory, the single-sided representation of the elastodynamic homogeneous Green's function is accurate for waves that are non-evanescent at the virtual source and receiver depth levels. In our first numerical experiment, we only consider propagating waves and confirm the theory within numerical precision. In our second experiment, we investigate wavefield components that are evanescent inside a thin layer, but propagating at the virtual source and receiver depth levels. In this case, the numerical result for the elastodynamic homogeneous Green's function using the single-sided representation performs well also for evanescent waves, except for a small relative error. The latter experiment is a tunnelling experiment, which performs better if the tunnelled layer is thin with respect to the inverse of the absolute vertical wavenumber. Although evanescent wave tunnelling appears to be possible in theory, in practice it will suffer from noise and it will not be possible to retrieve the evanescent components of the focusing function via the Marchenko method.

### 4.2. Conclusion

We presented a numerical example demonstrating that single-sided access to a medium suffices to correctly retrieve the non-evanescent components of an elastodynamic homogeneous Green's function with virtual sources and virtual receivers inside the medium. Despite the single-sided access to the medium, all events of the homogeneous Green's function including primaries, internal multiples and converted waves are represented correctly. Waves that are evanescent, either on the recording boundary, or at the virtual source/receiver depth level, are neglected by the single-sided representation of the elastodynamic homogeneous Green's function, and hence, can lead to artefacts. Nevertheless, the evanescent components can be filtered to suppress these artefacts. The remaining, i.e. propagating, components can be used e.g. for imaging.

In addition, we observed in a numerical experiment that the elastodynamic single-sided homogeneous Green's function representation can tunnel evanescent components through a thin layer. In theory, this tunnelling of evanescent waves is independent of the thickness of the tunnelled layer. In practice, the single-sided representation of tunnelled waves has limited numerical accuracy because it has to compensate for the exponential amplitude decay of evanescent waves, which is stronger for thicker layers. Hence, the tunnelled layer has to be sufficiently thin with respect to the inverse of the absolute vertical wavenumber of the evanescent wave. Note that we refer to tunnelling because the single-sided representation requires that at the virtual source and receiver depth levels the elastodynamic homogeneous Green's function is propagating.

To conclude, we provided a mathematical framework to create virtual wavefield measurements inside a medium that is only accessible from a single side. We foresee potential applications in the fields of imaging, time-lapse monitoring, forecasting of the medium response as well as localisation of passive sources.

**Acknowledgements**

We thank Carlos Alberto da Costa Filho and one anonymous reviewer for their constructive comments that helped us to improve the paper. Further, we would like to acknowledge the Waves project for offering such a great research network. Last but not least, we are grateful to our co-workers for all the inspiring discussions. This work was supported by the European Union's Horizon 2020 research and innovation programme: Marie Skłodowska-Curie [grant number 641943] and European Research Council [grant number 742703].

**Appendix A. Elastodynamic single-sided homogeneous Green's function representation**

In the following, we summarise the derivation of the elastodynamic single-sided homogeneous Green's function representation [6]. First, we show the decomposed reciprocity theorems and introduce the two states that are used in the derivation. Second, we explain the properties of the focusing function. Third, we derive the elastodynamic single-sided homogeneous Green's function representation by inserting the focusing function and the Green's function in the decomposed reciprocity theorems.

*A.1. Reciprocity theorems and the two states*

We define a domain $\mathbb{D}$ with infinite horizontal extent. The $x_3$ axis is defined along the depth direction as downward pointing. The domain is enclosed by $\partial\mathbb{D}_0$ ($x_3 = x_{3,0}$) at the top, $\partial\mathbb{D}_r$ ($x_3 = x_{3,r}$) at the bottom and a cylindrical boundary $\partial\mathbb{D}_{cyl}$ with infinite radius at the side. Besides, we introduce two wavefield states A and B with independent decomposed wavefield vectors $\mathbf{p}_{A,B}$ as well as independent decomposed source vectors $\mathbf{s}_{A,B}$, both in the space–frequency domain. We assume the medium in which these wavefields and sources exist is lossless, and that the medium parameters of the states A and B are identical inside the domain $\mathbb{D}$. Now we can formulate the reciprocity theorems of the convolution- and correlation-type,

$$\int_{\partial\mathbb{D}} \mathbf{p}_A^T \mathbf{N} \mathbf{p}_B n_3 \mathrm{d}^2\mathbf{x} = \int_{\mathbb{D}} \left( \mathbf{s}_A^T \mathbf{N} \mathbf{p}_B + \mathbf{p}_A^T \mathbf{N} \mathbf{s}_B \right) \mathrm{d}^3\mathbf{x}, \tag{A.1}$$

$$\int_{\partial\mathbb{D}} \mathbf{p}_A^\dagger \mathbf{J} \mathbf{p}_B n_3 \mathrm{d}^2\mathbf{x} = \int_{\mathbb{D}} \left( \mathbf{s}_A^\dagger \mathbf{J} \mathbf{p}_B + \mathbf{p}_A^\dagger \mathbf{J} \mathbf{s}_B \right) \mathrm{d}^3\mathbf{x}. \tag{A.2}$$

Here, the dagger symbol † denotes transposition and complex-conjugation. On the boundary $\partial\mathbb{D}$, the derivation of the reciprocity theorem of the correlation-type (Eq. (A.2)) uses the symmetry relation $\mathcal{L}^\dagger \mathbf{K} = \mathbf{J}\mathcal{L}^{-1}$, which only holds for propagating waves [28]. Note that $\mathcal{L} = \mathcal{L}(\mathbf{x}, \omega)$ is the operator matrix that composes wavefields. Consequently, we neglect evanescent waves on the boundary $\partial\mathbb{D}$. The boundary $\partial\mathbb{D}$ consists of the top and bottom boundaries of the domain, $\partial\mathbb{D}_0 \cup \partial\mathbb{D}_r$. Integral contributions over the cylindrical boundary $\partial\mathbb{D}_{cyl}$ vanish (if the medium has infinite horizontal extent) because the integrand is proportional to one divided by the radius squared ($\propto \frac{1}{R^2}$) whereas the cylindrical surface is proportional to the radius ($\propto R$). In the boundary integrals, $n_3$ is the $x_3$ component of the outward-pointing normal vector on the boundary $\partial\mathbb{D}$ ($n_3 = -1$ on $\partial\mathbb{D}_0$, $n_3 = +1$ on $\partial\mathbb{D}_r$).

The single-sided homogeneous Green's function representation is derived by evaluating the two reciprocity theorems using a specific definition of states A and B. First, we define state B in the actual medium. Let the upper boundary $\partial\mathbb{D}_0$ be at $x_3 = x_{3,0}$. Above $\partial\mathbb{D}_0$, i.e. $x_3 \leq x_{3,0}$, the state B medium is reflection-free. Below $\partial\mathbb{D}_0$, i.e. $x_3 > x_{3,0}$, the actual medium is inhomogeneous and elastic. We define the state B wavefield to be the medium's Green's function $\mathbf{\Gamma}(\mathbf{x}, \mathbf{x}_s, \omega)$ related to a source at $\mathbf{x}_s$ with $x_{3,s} > x_{3,0}$, and a receiver at $\mathbf{x}$,

$$\mathbf{p}_B = \mathbf{\Gamma}(\mathbf{x}, \mathbf{x}_s, \omega), \quad \mathbf{s}_B = \mathbf{I}\delta(\mathbf{x} - \mathbf{x}_s). \tag{A.3}$$

Second, we define state A. We choose a point $\mathbf{x}_r$ below $\partial\mathbb{D}_0$. The state A medium is defined equal to the state B medium for $x_{3,0} \leq x_3 \leq x_{3,r}$, and for $x_3 \geq x_{3,r}$ the state A medium is reflection-free. Often, the state A medium is referred to as the so-called truncated medium. The state A wavefield is defined as the focusing function $\mathbf{F}(\mathbf{x}, \mathbf{x}_r, \omega)$. By definition the source term of the focusing function is zero,

$$\mathbf{p}_A = \mathbf{F}(\mathbf{x}, \mathbf{x}_r, \omega), \quad \mathbf{s}_A = \mathbf{0}. \tag{A.4}$$

### A.2. Focusing function

The downgoing part of the focusing function $\mathbf{F}^+(\mathbf{x}, \mathbf{x}_r, \omega)$ is the inverse of a transmission response related to sources at the boundary $\partial \mathbb{D}_0$ and a receiver at $\mathbf{x}_r$,

$$\int_{\partial \mathbb{D}_0} \mathbf{T}^+(\mathbf{x}, \mathbf{x}', \omega) \mathbf{F}^+(\mathbf{x}', \mathbf{x}_r, \omega) d^2 \mathbf{x}'_H \bigg|_{x_3 = x_{3,r}} = \mathbf{I} \, \delta(\mathbf{x}_H - \mathbf{x}_{H,r}), \tag{A.5}$$

and it obeys the focusing condition,

$$\mathbf{F}(\mathbf{x}, \mathbf{x}_r, \omega)|_{x_3 = x_{3,r}} = \mathbf{I}_1 \, \delta(\mathbf{x}_H - \mathbf{x}_{H,r}), \tag{A.6}$$

with $\mathbf{F}$ containing $\mathbf{F}^+$ and $\mathbf{F}^-$, see Eq. (16). The upgoing focusing function $\mathbf{F}^-(\mathbf{x}, \mathbf{x}_r, \omega)$ is the reflection response of the downgoing focusing function in the truncated medium.

In a physical interpretation the focusing function, when transformed to the time domain, is a wavefield injected at $\partial \mathbb{D}_0$, which focuses at time zero at the location $\mathbf{x}_r$. Fig. 1(b) depicts the focusing function in a cartoon. The solid arrows represent the downgoing focusing function $\mathbf{F}^+$. When the downgoing focusing function is sent into the truncated medium it is partially reflected, leading to the upgoing focusing function, indicated by dashed arrows in Fig. 1(b). In the absence of a coda, the upgoing focusing function would be reflected downward again. Consequently, the focusing function would not focus at $\mathbf{x}_r$. This scenario is prevented by sending a coda of the downgoing focusing function into the medium to cancel the downward reflections of the upgoing focusing function. The coda is also shown in Fig. 1(b).

For a 3D acoustic medium, the focusing function can be retrieved from the reflection response of the medium combined with a smooth velocity model via the Marchenko method [e.g.7–9]. For an elastic medium, the focusing function retrieval still requires additional information about the medium [11].

### A.3. Derivation

We insert states A and state B (Eqs. (A.3), (A.4)) in the reciprocity theorems of the convolution- and the correlation-type (Eqs. (A.1), (A.2)) and evaluate them in the domain $\mathbb{D}_r$ bounded by $\partial \mathbb{D}_0$ at the top and by $\partial \mathbb{D}_r$ at the bottom. Note that the state A and state B media are identical in the domain $\mathbb{D}_r$, which is a necessary condition for Eqs. (A.1)–(A.2). Using the focusing condition of Eq. (A.6), we find,

$$\mathbf{I}_1^T \mathbf{N} \mathbf{\Gamma}(\mathbf{x}_r, \mathbf{x}_s, \omega) - H(x_{3,r} - x_{3,s}) \mathbf{F}^T(\mathbf{x}_s, \mathbf{x}_r, \omega) \mathbf{N} = \int_{\partial \mathbb{D}_0} \mathbf{F}^T(\mathbf{x}, \mathbf{x}_r, \omega) \mathbf{N} \mathbf{\Gamma}(\mathbf{x}, \mathbf{x}_s, \omega) \, d^2 \mathbf{x}_H, \tag{A.7}$$

and,

$$\mathbf{I}_1^T \mathbf{J} \mathbf{\Gamma}^*(\mathbf{x}_r, \mathbf{x}_s, \omega) - H(x_{3,r} - x_{3,s}) \mathbf{F}^T(\mathbf{x}_s, \mathbf{x}_r, \omega) \mathbf{J} = \int_{\partial \mathbb{D}_0} \mathbf{F}^T(\mathbf{x}, \mathbf{x}_r, \omega) \mathbf{J} \mathbf{\Gamma}^*(\mathbf{x}, \mathbf{x}_s, \omega) \, d^2 \mathbf{x}_H. \tag{A.8}$$

$H(x_3)$ is the Heaviside function. We multiply Eq. (A.8) by $\mathbf{K}$ from the right and substitute the identities $\mathbf{J} = \mathbf{N}\mathbf{K}$ as well as $\mathbf{J}\mathbf{K} = \mathbf{N}$,

$$\mathbf{I}_1^T \mathbf{N} \mathbf{K} \mathbf{\Gamma}^*(\mathbf{x}_r, \mathbf{x}_s, \omega) \mathbf{K} - H(x_{3,r} - x_{3,s}) \mathbf{F}^T(\mathbf{x}_s, \mathbf{x}_r, \omega) \mathbf{N} = \int_{\partial \mathbb{D}_0} \mathbf{F}^T(\mathbf{x}, \mathbf{x}_r, \omega) \mathbf{N} \mathbf{K} \mathbf{\Gamma}^*(\mathbf{x}, \mathbf{x}_s, \omega) \mathbf{K} \, d^2 \mathbf{x}_H. \tag{A.9}$$

We eliminate the term with the Heaviside function by subtracting Eq. (A.9) from Eq. (A.7). The resulting expression can be written in terms of the homogeneous Green's function (using Eq. (11)),

$$\mathbf{I}_1^T \mathbf{N} \mathbf{\Gamma}_h(\mathbf{x}_r, \mathbf{x}_s, \omega) = \int_{\partial \mathbb{D}_0} \mathbf{F}^T(\mathbf{x}, \mathbf{x}_r, \omega) \mathbf{N} \mathbf{\Gamma}_h(\mathbf{x}, \mathbf{x}_s, \omega) \, d^2 \mathbf{x}_H. \tag{A.10}$$

The multiplication by $\mathbf{I}_1^T \mathbf{N}$ from the left in Eq. (A.10) deletes the upper submatrices of the homogeneous Green's function $\mathbf{\Gamma}_h(\mathbf{x}_r, \mathbf{x}_s, \omega)$. We retrieve the complete matrix $\mathbf{\Gamma}_h(\mathbf{x}_r, \mathbf{x}_s, \omega)$ by multiplying Eq. (A.10) by $\mathbf{I}_2$ from the left,

$$\mathbf{\Gamma}_2(\mathbf{x}_r, \mathbf{x}_s, \omega) = \mathbf{I}_2 \mathbf{I}_1^T \mathbf{N} \mathbf{\Gamma}_h(\mathbf{x}_r, \mathbf{x}_s, \omega) = \begin{pmatrix} \mathbf{O} & \mathbf{O} \\ \mathbf{G}^{-,+} - (\mathbf{G}^{+,-})^* & \mathbf{G}^{-,-} - (\mathbf{G}^{+,+})^* \end{pmatrix}, \tag{A.11}$$

and by using Eqs. (11) and (12),

$$\mathbf{\Gamma}_h(\mathbf{x}_r, \mathbf{x}_s, \omega) = \mathbf{\Gamma}_2(\mathbf{x}_r, \mathbf{x}_s, \omega) - \mathbf{K} \mathbf{\Gamma}_2^*(\mathbf{x}_r, \mathbf{x}_s, \omega) \mathbf{K}. \tag{A.12}$$

From Eqs. (A.10) and (A.11) it follows that the Green's function $\mathbf{\Gamma}_2(\mathbf{x}_r, \mathbf{x}_s, \omega)$ is defined as,

$$\mathbf{\Gamma}_2(\mathbf{x}_r, \mathbf{x}_s, \omega) = \int_{\partial \mathbb{D}_0} \mathbf{I}_2 \mathbf{F}^T(\mathbf{x}, \mathbf{x}_r, \omega) \mathbf{N} \mathbf{\Gamma}_h(\mathbf{x}, \mathbf{x}_s, \omega) d^2 \mathbf{x}. \tag{A.13}$$

Eqs. (A.12)–(A.13) together form the single-sided homogeneous Green's function representation for $\mathbf{\Gamma}_h(\mathbf{x}_r, \mathbf{x}_s, \omega)$. The right-hand side of Eq. (A.13) contains the homogeneous Green's function $\mathbf{\Gamma}_h(\mathbf{x}, \mathbf{x}_s, \omega)$, for which we can obtain a single-sided representation in a similar way. First, in Eq. (A.10), we substitute $\mathbf{x}$ by $\mathbf{x}'$ on $\partial\mathbb{D}_0'$ (just above $\partial\mathbb{D}_0$), $\mathbf{x}_s$ by $\mathbf{x}$ (on $\partial\mathbb{D}_0$) and $\mathbf{x}_r$ by $\mathbf{x}_s$,

$$\mathbf{I}_1^T \mathbf{N}\mathbf{\Gamma}_h(\mathbf{x}_s, \mathbf{x}, \omega) = \int_{\partial\mathbb{D}_0'} \mathbf{F}^T(\mathbf{x}', \mathbf{x}_s, \omega) \mathbf{N}\mathbf{\Gamma}_h(\mathbf{x}', \mathbf{x}, \omega)\, d^2\mathbf{x}_H'. \quad (A.14)$$

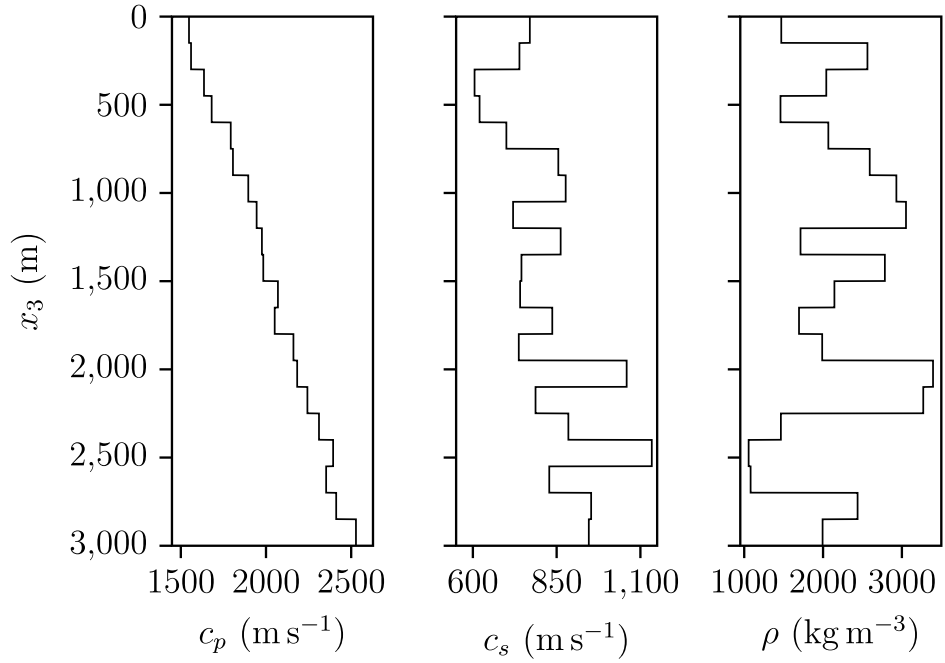

**Fig. B.11.** Layered model. The model depth ranges from 0 m to 3000 m, the lateral distance ranges from −12 812.5 m to 12 800 m. The P-wave velocity, S-wave velocity and density are denoted by $c_p$, $c_s$ and $\rho$, respectively.

Second, we multiply Eq. (A.14) by $\mathbf{N}$ from the right, transpose the result and apply source–receiver reciprocity ($\mathbf{N}\mathbf{\Gamma}_h^T(\mathbf{x}_s, \mathbf{x}, \omega)\mathbf{N} = \mathbf{\Gamma}_h(\mathbf{x}, \mathbf{x}_s, \omega)$),

$$\mathbf{\Gamma}_h(\mathbf{x}, \mathbf{x}_s, \omega)\mathbf{I}_1 = \int_{\partial\mathbb{D}_0'} \mathbf{\Gamma}_h(\mathbf{x}, \mathbf{x}', \omega)\mathbf{F}(\mathbf{x}', \mathbf{x}_s, \omega)\, d^2\mathbf{x}_H'. \quad (A.15)$$

Multiplication by matrix $\mathbf{I}_1$ deletes part of the homogeneous Green's function $\mathbf{\Gamma}_h(\mathbf{x}, \mathbf{x}_s, \omega)$. The full matrix $\mathbf{\Gamma}_h(\mathbf{x}, \mathbf{x}_s, \omega)$ is constructed by multiplying by $\mathbf{I}_1^T$ from the right,

$$\mathbf{\Gamma}_1(\mathbf{x}, \mathbf{x}_s, \omega) = \mathbf{\Gamma}_h(\mathbf{x}, \mathbf{x}_s, \omega)\mathbf{I}_1\mathbf{I}_1^T = \begin{pmatrix} \mathbf{G}^{+,+} - (\mathbf{G}^{-,-})^* & \mathbf{O} \\ \mathbf{G}^{-,+} - (\mathbf{G}^{+,-})^* & \mathbf{O} \end{pmatrix}, \quad (A.16)$$

and by using the definition of the homogeneous Green's function (Eqs. (11), (12)),

$$\mathbf{\Gamma}_h(\mathbf{x}, \mathbf{x}_s, \omega) = \mathbf{\Gamma}_1(\mathbf{x}, \mathbf{x}_s, \omega) - \mathbf{K}\mathbf{\Gamma}_1^*(\mathbf{x}, \mathbf{x}_s, \omega)\mathbf{K}, \quad (A.17)$$

where $\mathbf{\Gamma}_1(\mathbf{x}, \mathbf{x}_s, \omega)$ is defined as,

$$\mathbf{\Gamma}_1(\mathbf{x}, \mathbf{x}_s, \omega) = \int_{\partial\mathbb{D}_0'} \mathbf{\Gamma}_h(\mathbf{x}, \mathbf{x}', \omega)\mathbf{F}(\mathbf{x}', \mathbf{x}_s, \omega)\mathbf{I}_1^T d^2\mathbf{x}'. \quad (A.18)$$

Eqs. (A.17)–(A.18) together form the single-sided homogeneous Green's function representation for $\mathbf{\Gamma}_h(\mathbf{x}, \mathbf{x}_s, \omega)$.

In summary, we derived a single-sided representation of the homogeneous Green's function $\mathbf{\Gamma}_h(\mathbf{x}_r, \mathbf{x}_s, \omega)$ consisting of two steps. In the first step (Eqs. (A.17)–(A.18)) a virtual source is created inside the medium. In the second step (Eqs. (A.12)–(A.13)) a virtual receiver is created inside the medium.

### Appendix B. 20-layer model

The numerical experiment of Section 3.2 is repeated for the 20 layer model shown in Fig. B.11.

We model the reflection response and the required focusing functions to create a virtual source at $\mathbf{x}_s = (0\,\text{m}, 1500\,\text{m})^T$ and virtual receivers on a grid with a depth range from 0 m to 3000 m and a lateral distance range from −2000 m to 2000 m. The spatial sampling interval is 12.5 m in both the vertical and horizontal direction.

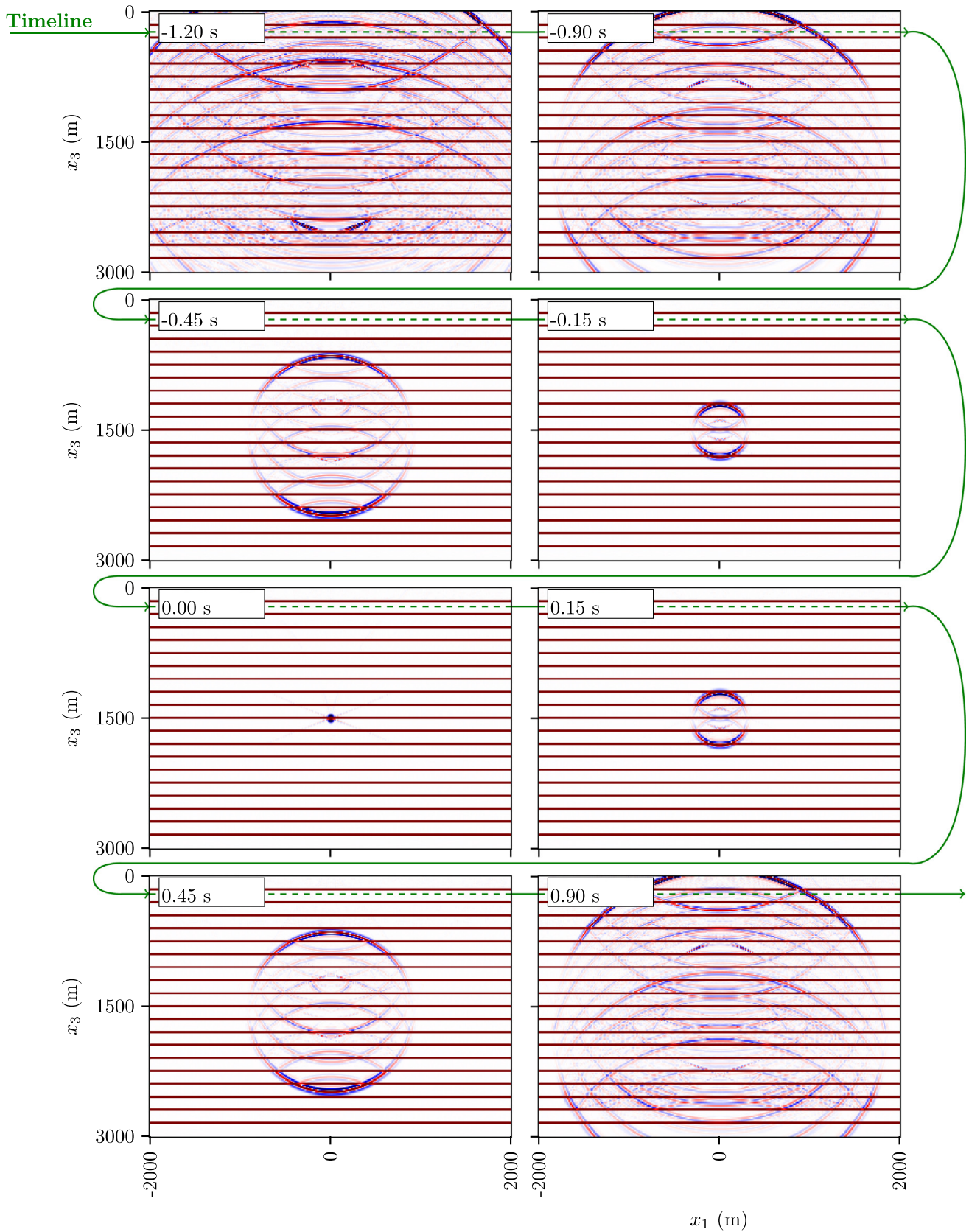

**Fig. B.12.** Single-sided representation of the elastodynamic homogeneous Green's function. The time slices show the result of the elastodynamic single-sided homogeneous Green's function representation $\mathbf{G}_h^{v,f}(\mathbf{x}_r, \mathbf{x}_s, t)$ related to virtual source ($f_3$) at $\mathbf{x}_s = (0\,\text{m}, 1500\,\text{m})^T$ and virtual receivers ($v_3$) placed on a grid with a depth range from 0 m to 3000 m and a lateral distance range from −2000 m to 2000 m. The spatial sampling interval is 12.5 m in both horizontal and vertical direction. The time slices were multiplied by a gain function ($\times e^{1.5|t|}$) to emphasise the late arrivals.

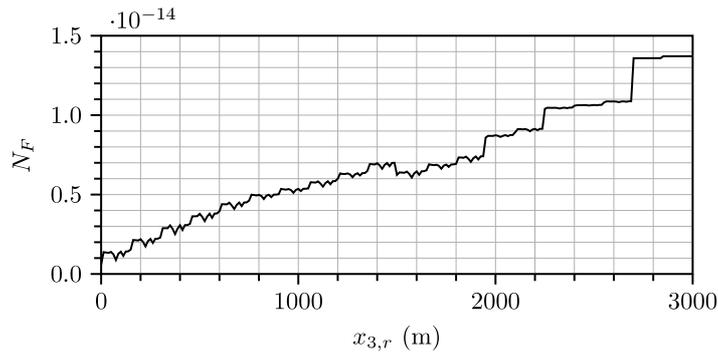

**Fig. B.13.** Error analysis. Normalised Frobenius norm $N_F$ of the difference $\Delta\tilde{\Gamma}_h(k_1, x_{3,r}, x_{3,s}, \omega)$ between the elastodynamic single-sided homogeneous Green's function representation (see Fig. B.12) and its modelled equivalent as a function of virtual receiver depth $x_{3,r}$.

From the reflection response and the focusing function we compute the single-sided representation of the elastodynamic homogeneous Green's function $\tilde{\Gamma}_h(k_1, x_{3,r}, x_{3,s}, \omega)$ and apply a $k_1$–$\omega$ filter (determined by the P-wave velocity as a function of the virtual receiver depth $x_{3,r}$). We compose the result according to Eq. (31) and obtain the full elastodynamic homogeneous Green's function $\tilde{\mathbf{G}}_h(k_1, x_{3,r}, x_{3,s}, \omega)$. Next, we apply a transformation to the space–time domain and a convolution with a 30 Hz Ricker wavelet. Fig. B.12 displays the $(v_3, f_3)$ component of the resulting elastodynamic homogeneous Green's function $\mathbf{G}_h^{v,f}(\mathbf{x}_r, \mathbf{x}_s, t)$.

To analyse the accuracy of the single-sided representation, we model the elastodynamic homogeneous Green's function for an actual source at $\mathbf{x}_s = (0\,\mathrm{m}, 1500\,\mathrm{m})^T$. We compute the difference between the modelled and the single-sided representation of the elastodynamic homogeneous Green's function. To exclude the evanescent wavefield, we element-wise multiply the residual $\Delta\tilde{\Gamma}_h(k_1, x_{3,r}, x_{3,s}, \omega)$ by a $k_1$–$\omega$ filter $\tilde{\mathbf{M}}$, which is determined by the maximum propagation velocity of the medium. Subsequently, we evaluate the normalised Frobenius norm $N_F = \frac{1}{\sqrt{4nt\,4nr}}\|\tilde{\mathbf{M}} \circ \Delta\tilde{\Gamma}_h(k_1, x_{3,r}, x_{3,s}, \omega)\|_2$ and show the result as a function of virtual receiver depth $x_{3,r}$ in Fig. B.13. The error plot demonstrates that, also in case of the 20 layer model, the single-sided homogeneous Green's function representation is accurate for propagating waves within numerical precision.